\pdfoutput=1
\documentclass[12pt]{article}
\usepackage[utf8]{inputenc}
\usepackage[T1]{fontenc}
\usepackage{hyperref}
\usepackage{amsmath}
\usepackage{amssymb}
\DeclareMathOperator{\ad}{\mathop{ad}}
\DeclareMathOperator{\trh}{\mathop{\hat{tr}}}
\newcommand{\normord}[1]{\mathopen{:}\mathinner{#1}\mathclose{:}}
\usepackage[margin=3cm]{geometry}
\usepackage{lipsum}
\usepackage{breqn}
\usepackage{tikz}
\usepackage{float}
\usepackage[absolute]{textpos}
\usepackage{pgfplots}
\usetikzlibrary{shapes.multipart,calc,datavisualization}
\DeclareMathOperator*{\tr}{\mathop{tr}}

\usepackage{empheq}

\usepackage{amsmath}
\usepackage{amssymb}

\ifcsname endofdump\endcsname\endofdump\fi

\numberwithin{equation}{section}
\usepackage{cleveref}
\usepackage[style=trad-alpha,maxalphanames=4, minalphanames=3, url=false]{biblatex}
\usepackage{authblk}
\author{Yaroslav Drachov\thanks{\href{mailto:drachov.yai@phystech.edu}{\texttt{drachov.yai@phystech.edu}}}}
\affil{Moscow Institute of Physics and Technology,\authorcr 141701 Dolgoprudny, Russia}
\date{}
\title{Generalized \(\widetilde{W}\) algebras}
\hypersetup{
 pdfauthor={},
 pdftitle={Generalized \(\widetilde{W}\) algebras},
 pdfkeywords={},
 pdfsubject={},
 pdfcreator={Emacs 29.3 (Org mode 9.7-pre)}, 
 pdflang={English}}
\begin{document}

\maketitle
\begin{textblock}{5}(13.59,2.65)
\begin{flushright}
MIPT/TH-12/24
\end{flushright}
\end{textblock} 
\begin{abstract}
Recently, a new generalized family of infinite-dimensional \( \widetilde{W}  \) algebras, each associated with a particular element of a commutative subalgebra of the \( W_{1+\infty} \) algebra, was described. This paper provides a comprehensive account of the aforementioned association, accompanied by the requisite proofs and illustrative examples. This approach allows a derivation of Ward identities for selected WLZZ matrix models and the expansion of corresponding \( W \)-operators  in terms of an infinite set of variables \( p_k \).
\end{abstract}
\tableofcontents
\section{Introduction}
\label{sec:org83d5c97}
The \(\widetilde{W}\)-algebras first appeared in \cite{marshakov-1992-from-viras,ahn-1992-one-point} in the context of constraints on tau functions of two-matrix models. Here, the first series of such algebras was discovered; they were called \(\widetilde{W}^{(n)}\) algebras. After a period of time, it was discovered that in the context of the generalized Kontsevich model, there is also a second series of a similar nature, designated as a \(\widetilde{W}^{(+,n) }\) series \cite{mironov-1996-unitar-matrix}. Two of these algebras have since been referred to as \(\widetilde{W}^{(\pm,n)}\) algebras. The simplest \(\widetilde{W}^{(\pm,n)}\) algebra is nothing but the Borel subalgebra of the Virasoro algebra. The higher spin algebras are no longer Lie algebras and can be described by commutation relations \cite{marshakov-1992-from-viras,mironov-1996-unitar-matrix}. Additionally, there are alternative definitions of such algebras. We are particularly interested in their realization as operators acting on the space of scalar functions of matrix variables \(\Lambda\). These functions can be thought of as functions of (possibly infinite-dimensional) vectors \(\mathbf{p}\) with the entities \(p_k = \tr \Lambda^k\).

A new discovery regarding the subject of the \(\widetilde{W}\) algebras was made in the works \cite{alexandrov-2014-kp-integ,mironov-2023-spect-curves}. In the former, a new series emerged that is intricately linked with the concept of \(\widetilde{W}^{(m,+)}\). However, in the latter, some indications of the generalization of these three series were uncovered. This was made possible by the recent advances \cite{mironov-2023-spect-curves,mironov-2023-kp-integ,mironov-2023-inter-matrix,mironov-2023-commut-subal,mironov-2023-many-body,mironov-2023-commut-famil,drachov-2024-w-w} in the study of the WLZZ models \cite{wang-2022-super-partit,wang-2022-cft-approac} and their connection to the \(W_{1+\infty}\) algebra  \cite{pope-1990-ideal-kac,pope-1990-w-racah,pope-1990-new-higher,kac-1993-quasif-highes,miki-2007-analog-w-algeb,fukuma-1992-infin-dimen,bakas-1992-beyon-large,kac-1996-repres-theor,pope-1990-compl-struc,awata-1995-repres-theor,bakas-1993-logar-deriv,frenkel-1995-with-centr-charg}.

In particular, it was observed in \cite{mironov-2023-spect-curves} that distinguished sums of the \(\widetilde{W}\) operators give elements of some commutative subalgebras of the \(W_{1+\infty}\) (strictly speaking, their matrix representation), namely
\begin{equation}
\label{eq:57}
H_{-n}^{(-1)} = \tr \left( \frac{\partial }{\partial \Lambda}  \right)^n = \sum_{k}^{} p_k \widetilde{W}_{k+n}^{(-,n)},    
\end{equation}
\begin{equation}
\label{eq:50}
H_n^{(1)} = \tr \left( \Lambda \frac{\partial }{\partial \Lambda} \Lambda \right)^n = \sum_{k}^{} p_k \widetilde{W}_{k-n}^{(+,n)},
\end{equation}
whereas the commutative subalgebras of the \(W_{1+\infty}\) enumerated by upper index are given as
\begin{equation}
\label{eq:65}
H_{n }^{(m)} = \tr \left( \left( \Lambda \frac{\partial }{\partial \Lambda}  \right)^m \Lambda \right)^{n}, \qquad H_{-n}^{(-m)} = \tr \left(\Lambda^{-1} \left( \Lambda \frac{\partial }{\partial \Lambda}  \right)^m  \right)^{n}.
\end{equation}
for both \(n\) and \(m\) positive. The question of generalizing the \(\widetilde{W}\) operators to be associated with not only two marked commutative subalgebras of the \(W_{1+\infty}\) \eqref{eq:57}, \eqref{eq:50}, but with the entire set of them \eqref{eq:65}, is worthy of consideration. To be precise, the question is about identifying \(\widetilde{W}^{(\pm m,\pm n)}_k\) operators such that
\begin{equation}
\label{eq:88}
H_{-n}^{(-m)} = \tr \left(\Lambda^{-1} \left( \Lambda \frac{\partial }{\partial \Lambda}  \right)^m  \right)^{n} =
\sum_{k }^{} p_k \widetilde{W}_{k + n}^{(-m, -n)},
\end{equation}
\begin{equation}
\label{eq:62}
H_n^{(m)}= \tr \left( \left( \Lambda \frac{\partial }{\partial \Lambda}  \right)^m \Lambda \right)^{n} = \sum_{k}^{}p_k \widetilde{W}_{k-n}^{(m,n)}.
\end{equation}
In this work, we address and solve this problem with the main result being the recursive definition for the \(\widetilde{W}^{(\pm m,\pm n)}_k\) operators (see eqs.  \eqref{eq:185}, \eqref{eq:200}, \eqref{eq:251}). It is also of significance that the Ward identities for the WLZZ models (eq. \eqref{eq:64}) are explicitly found.

\begin{figure}[h]
\centering
\begin{tikzpicture}
    \tikzset{ 
        every pin/.style={fill=gray!30!white,rectangle,rounded corners=3pt,font=\tiny,pin distance=5pt},
        small dot/.style={fill,circle,scale=0.3}, 
    }
\draw[->, thick] (-7.5,0) -- (7.5,0) node[above left] {\( n \)};
\draw[->, thick] (0,0) -- (0,5);
\draw[ font=\small,ultra thick] (0,0) -- (7,3.5) node[above left] {\( \boldsymbol{m=1} \)};
\draw[ font=\small,ultra thick] (0,0) -- (-7,3.5) node[above right] {\( \boldsymbol{m=-1} \)};
\draw[ font=\small,thick] (0,0) -- (4.5,4.5) node[above] {\( m=2 \)};
\draw[ font=\small,thick] (0,0) -- (-4.5,4.5) node[above] {\( m=-2 \)};
\draw[ font=\small,thick] (0,0) -- (-3,4.5) node[above] {\( m=-3 \)};
\draw[ font=\small,thick] (0,0) -- (-2.25,4.5) node[above right] {\( m=-4 \)};
\draw[ font=\small,thick] (0,0) -- (2.25,4.5) node[above left] {\( m=4 \)};
\draw[ font=\small,thick] (0,0) -- (3,4.5) node[above] {\( m=3 \)}; 
\foreach \y in {0,...,4}      
{
  \node [small dot,pin=210:{\(H_{-1}^{(-\y )}= F_\y \)}]  () at (-2,\y) {};
  \node [small dot,pin=-30:{\(H_{1}^{(\y )}= E_\y \)}]  () at (2,\y) {};
}

\node [small dot,pin=-30:{\(L_0= H^{(1)}_0 \)}]  (0) at (0,1) {};
\node [small dot,pin=-30:{\(6W_0= H^{(2)}_0 \)}]  (0) at (0,2) {};
\node [small dot,pin=-30:{\(H^{(3)}_0 \)}]  (0) at (0,3) {};
\node [small dot,pin=-30:{\(H^{(4)}_0 \)}]  (0) at (0,4) {};
\node [small dot,pin=-30:{\(H_2^{(1)} \)}]  (0) at (4,2) {};
\node [small dot,pin=-30:{\(H_2^{(2)} \)}]  (0) at (4,4) {};
\node [small dot,pin=-30:{\(H_3^{(1)} \)}]  (0) at (6,3) {}; 
\node [small dot,pin=210:{\(H_{-2}^{(-1)} \)}]  (0) at (-4,2) {};
\node [small dot,pin=210:{\(H_{-2}^{(-2)} \)}]  (0) at (-4,4) {}; 
\node [small dot,pin=210:{\(H_{-3}^{(-1)} \)}]  (0) at (-6,3) {};
\foreach \x in {-3,...,3}
  \foreach \y in {0,...,4}  
     \node [small dot] () at (2*\x,\y) {};  
\end{tikzpicture} 
\caption[]{Commutative subalgebras (integer rays) of \( W_{1+\infty} \) algebra depicted on a 2\( d \) lattice}  
\label{fig:1}   
\end{figure}
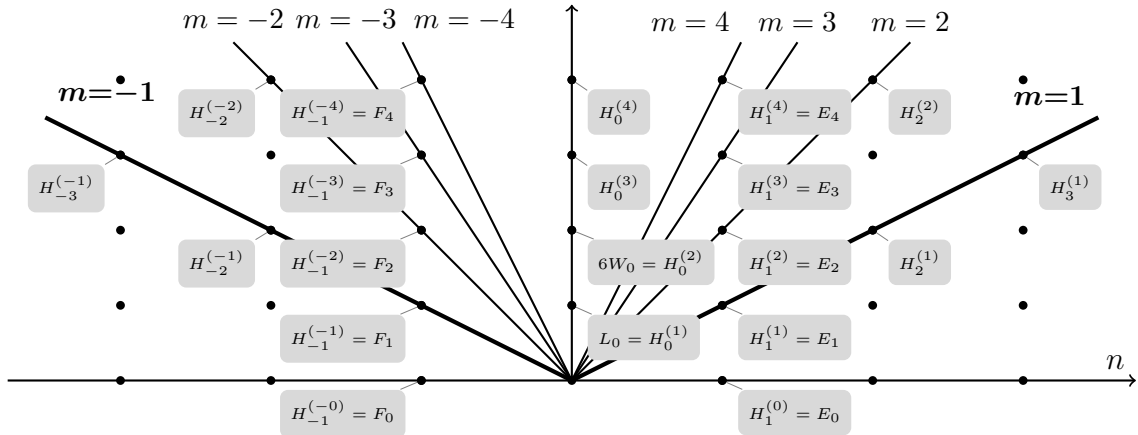

Some remarks regarding the \(W_{1+\infty}\) algebra are in order. \(W_{1+\infty}\) is a Lie algebra that can be generated by three operators: the so-called cut-and-join operator, denoted by \(W_0\), \(E_0\) and \(F_0\). Defining
\begin{equation}
\label{eq:69}
E_m = \ad_{W_0}^m E_0, \qquad F_m= \ad_{W_0}^m F_0,
\end{equation}
the elements of the discussed commutative subalgebras are given as
\begin{equation}
\label{eq:70}
H_n^{(m)} = \ad_{E_{m+1}}^{n-1} E_m, \qquad H_{-n}^{(-m)}= \ad_{F_{m+1}}^{n-1} F_m.
\end{equation}
All the described operators can be represented on a 2\(d\) lattice (fig. \ref{fig:1}).
It is evident that there should also be a commutative subalgebra constituted by elements of zero
grading, for instance the commutators of \(E_i\) and \(F_j\). It can be demonstrated that such a subalgebra is comprised of the elements
\begin{equation}
\label{eq:71}
H^{(i+j - 1)}_0 = \left[ F_i, E_j \right].
\end{equation}
The first three of these elements are distinguished, being simply the multiplication by \(N\), the Virasoro operator \(L_0\) and something proportional to the cut-and-join operator \(W_0\).
A detailed discussion of the operators \(H_0^{(m)}\) and the corresponding \(\widetilde{W}\) algebras
\begin{equation}
\label{eq:87}
H_0^{(m)} = \sum_{k}^{} p_k \widetilde{W}_k^{(m,0)}
\end{equation}
is also planned for this paper. The fundamental result is the discovery of a recursive definition for the \(\widetilde{W}_k^{(m,0)}\)-s, which revealed that the \(H_0^{(m)}\)-s are, in fact, Casimirs (see, for example, \eqref{eq:229}). This implies that the associated partition functions generate the Hurwitz numbers with completed cycles. Once again, the knowledge of the \(\widetilde{W}^{(m,0)}\)-s allows one to explicitly construct the corresponding Ward identities, namely, the constraints on these tau functions.

The paper is organized as follows. In the section  \ref{sec:org850b9ed},
we review the known results in the field, setting up a stage for a subsequent step. Then, in section \ref{sec:orgbac4350}, the existing framework is extended to encompass not only \(m = \pm 1\) commutative rays of \(W_{1+\infty}\), but all of them. Section \ref{sec:org6fde181} is dedicated to the derivation of several pivotal matrix identities. Subsequently, in section \ref{sec:org23a3d4a}, a subtle connection between the vertical ray of \(W_{1+\infty}\) and Hurwitz numbers with completed cycles is described. Finally, section \ref{sec:org5104969} contains concluding remarks and future prospects of this work.
\section{A novel perspective on the known \(\widetilde{W}\) algebras}
\label{sec:org850b9ed}
In this section, we introduce the notation and set a stage for a generalization of \(\widetilde{W}\) algebras by considering three known algebras of such a nature. The objective of this section is to establish a connection between the existing knowledge and future possibilities, informed by the insights based on that knowledge. In particular, this section introduces a novel technique for deriving explicit recursive definitions of the \(\widetilde{W}\) operators.  This technique allows for the derivation of the known recursive definitions of the \(\widetilde{W}^{(\pm,n)}_k\) operators. In the penultimate part of this section a new recursive definition of the  \(\widetilde{W}^{(m,+)}_k\) operators is derived by the same reasoning. The last part is devoted to introducing the new \(\widetilde{W}^{(m,-)}_k\) algebra using the described procedure.
\subsection{\(m = 1\) integer ray}
\label{sec:orgfb88e3b}
The extensive information about the \(\widetilde{W}^{(n)}\) algebras described here could be found in \cite{marshakov-1992-from-viras}. Here, we will only review the information necessary for the subsequent steps of our discussion.

The first appearance of the \(\widetilde{W}\) algebras in the context of two-matrix models \(Z_{n} = Z_{n} (\mathbf{p})\), \(n \ge 1,\) is noteworthy. The partition function for a simplest two-matrix model should contain a mixing term between two matrices, as well as individual potentials for each of them
\begin{equation}
\label{eq:113}
Z=\iint dX dY e^{\tr(W(Y) - XY + V(X))}.
\end{equation}
Here, integral is taken over \(N \times N\) Hermitian matrices \(X\) and \(Y\). The Ward identities for such models can be written in terms of the \(\widetilde{W}^{(n)}\) operators described below. However, for the sake of simplicity, let us consider a narrower subset of these models for which the potential \(V(X)\) is still taken to be an arbitrary analytic function, while \(W(Y)\) is restricted to be a monomial potential 
\begin{equation}
\label{eq:6}
Z_n = \iint dXdY \exp \tr \left( Y^n - YX + \sum_{k \ge 1}^{} \frac{p_k X^k}{k}\right)
\end{equation}
These partition functions possess what is referred to as a \(W\)-representation \cite{alexandrov-2023-w-operat,mironov-2023-inter-matrix,mironov-2023-kp-integ}.
\begin{equation}
\label{eq:21}
Z_{n} = e^{ \frac{1}{n} H_{n}^{(1)}} \cdot 1.
\end{equation}
They exhibit an infinite number of the infinitesimal symmetries
\begin{equation}
\label{eq:89}
Y \to Y + X^k,
\end{equation}
which correspond to the Ward identities
\begin{equation}
\label{eq:7}
\widetilde{W}_k^{(n)} Z_{n} = (k + n)
\frac{\partial Z_n}{\partial p_{k + n}}, \qquad k + n \ge 1.
\end{equation}
The \(\widetilde{W}^{(n)}\) algebra is by definition the algebra of constraints on a tau function of a particular multi-matrix model. In \cite{marshakov-1992-from-viras}, with the correction of \cite{mironov-2023-many-body}, the following definition of the \(\widetilde{W}^{(n)}_{k}\)-s  was formulated:
\begin{equation}
\label{eq:37}
\left( -\det\nolimits^{-N}\Lambda \frac{\partial }{\partial \Lambda^{-1}}\det\nolimits^{N}\Lambda  \right)^{n} f(\mathbf{p})= \sum_{k}^{}\Lambda^{k} \widetilde{W}_{k-n}^{(n)} f (\mathbf{p}) 
.\end{equation}
This was demonstrated to be equivalent to the following recursive definition:
\begin{equation}
\label{eq:42}
  \widetilde{W}_k^{(n + 1)} =   \sum_{m \ge 0}^{} p_m \widetilde{W}_{k + m}^{(n)} + \sum_{m = 1}^{k + n} m \frac{\partial }{\partial p_m}
  \widetilde{W}_{k - m}^{(n)} \qquad \text{for} \qquad k + n \ge 0.  
\end{equation}
\begin{equation}
\label{eq:121}
\widetilde{W}_k^{(n + 1)} = 0, \qquad \text{otherwise}.
\end{equation}
The base of this recursion is given as
\begin{equation}
\label{eq:48}
\widetilde{W}_k^{(0)}=\delta_{k,0}.
\end{equation}
In order to make everything crystal clear, let me list the first few \(\widetilde{W}\) operators of this series explicitly as a result of straightforward computation using the recursive relation (see Appendix \ref{sec:org1a24712}). For \(k \ge 0\), we have 
\begin{equation}
\label{eq:124}
\widetilde{W}_k^{(1)} = N \delta_{k,0} + k \frac{\partial }{\partial p_k}. 
\end{equation}
In turn, for \(k \ge -1\)
\begin{dmath}
\label{eq:125}
\widetilde{W}_k^{(2)} = N p_1 \delta_{k, -1} + N^2 \delta_{k,0} + \sum_{m \ge 0}^{}(k + m)p_m \frac{\partial }{\partial p_{k + m}} + Nk \frac{\partial }{\partial p_k} + \sum_{m = 1}^{k - 1} m (k - m) \frac{\partial ^2}{\partial p_m \partial p_{k - m}}.  
\end{dmath}
As previously noted, the \(\widetilde{W}^{(n)}_k\) operators satisfy eq. \eqref{eq:50}
\begin{equation}
\label{eq:90}
\tr \left( \Lambda \frac{\partial }{\partial \Lambda} \Lambda \right)^n = \sum_{k}^{} p_k \widetilde{W}_{k-n}^{(n)},
\end{equation}
The trace of \eqref{eq:37} reads as
\begin{equation}
\label{eq:91}
\tr \left( - \det\nolimits^{-N }\Lambda \frac{\partial }{\partial \Lambda^{-1}} \det\nolimits^{N}\Lambda  \right)^n =
\sum_{k}^{}p_k \widetilde{W}_{k-n}^{(n)},
\end{equation}
while acting on scalar functions of \(\Lambda\). Indeed, eq. \eqref{eq:50} was also written in this sense. 
One might inquire as to the veracity of the following relation
\begin{equation}
\label{eq:92}
\Lambda \frac{\partial }{\partial \Lambda} \Lambda = - \det\nolimits^{-N} \Lambda \frac{\partial }{\partial \Lambda^{-1}} \det\nolimits^N \Lambda ?
\end{equation}
The answer is affirmative; a proof can be found in Appendix \ref{sec:org8a93151}. In conclusion, we have established a non-recursive definition of the \(\widetilde{W}_k^{(n)}\)-s
\begin{equation}
\label{eq:95}
\left( \Lambda \frac{\partial }{\partial \Lambda} \Lambda \right)^n f(\mathbf{p})= \sum_{k}^{} \Lambda^k \widetilde{W}_{k-n}^{(n)} f(\mathbf{p}). 
\end{equation}
The subsequent question is whether it is possible to derive the recursive definition from the non-recursive one? Once more, the answer is affirmative, and it can be obtained through the proposed technique of explicit computation of matrix derivatives. We may now proceed. In order to obtain the base of the recursion, we must consider the case of \(n = 0\)
\begin{equation}
\label{eq:96}
\left( \Lambda \frac{\partial }{\partial \Lambda} \Lambda \right)^0 f(\mathbf{p}) = I f(\mathbf{p}) = \sum_{k}^{}\Lambda^k \widetilde{W}_{k}^{(0)} f(\mathbf{p}).
\end{equation}
It can be seen that this relation holds if and only if
\begin{equation}
\label{eq:97}
\widetilde{W}_k^{(0)}= \delta_{k,0}.
\end{equation}
Before evaluating the step of the recursion, it is possible to simplify the situation somewhat. 
Specifically, it is proposed that the left-hand side of \eqref{eq:95} is inspected. Initially, we have a smooth enough scalar function \(f\) from a matrix variable \(\Lambda\). Let us begin by examining the first few actions of the \(\Lambda \frac{\partial }{\partial \Lambda} \Lambda\) operator on \(f(\mathbf{p})\)
\begin{equation}
\label{eq:99}
\Lambda \frac{\partial }{\partial \Lambda} \Lambda f(\mathbf{p}) = \Lambda \left( \frac{\partial }{\partial \Lambda} \Lambda \right) f(\mathbf{p}) + \Lambda \left( \frac{\partial }{\partial \Lambda} f(\mathbf{p}) \right) \Lambda. 
\end{equation}
By utilizing the matrix calculus identities given in eqs. \eqref{eq:4} and \eqref{eq:5} we can get
\begin{dmath}
\label{eq:100}
\Lambda \frac{\partial }{\partial \Lambda}  \Lambda f(\mathbf{p}) =
N\Lambda f(\mathbf{p}) + \sum_{k \ge 1}^{}k \Lambda^{k+1} \frac{\partial }{\partial p_k} f(\mathbf{p}) =
\sum_{k\ge 1}^{} \Lambda^k\left( N \delta_{k-1,0} + (k-1) \frac{\partial }{\partial p_{k-1}}  \right) f(\mathbf{p}). 
\end{dmath}
Not only this type of reasoning allows one to obtain \(\widetilde{W}_k^{(1)}\), but it demonstrates that we indeed got the analytic matrix-valued function
\begin{equation}
\label{eq:101}
\Lambda \frac{\partial }{\partial \Lambda} \Lambda f(\mathbf{p}) = \sum_{k\ge 0}^{} \Lambda^k f_{k}(\mathbf{p}).  
\end{equation}
One can inquire whether this analytic matrix-valued function remains analytic matrix-valued throughout the subsequent steps of the evaluation process. This is a crucial question for us, as the existence of a lower bound of \(\Lambda\) powers in the expansion allows us to construct a recursion. In particular, it is proposed that the following expansion in non-negative powers of \(\Lambda\) exists
\begin{equation}
\label{eq:104}
\left( \Lambda \frac{\partial }{\partial \Lambda} \Lambda \right)^n f(\mathbf{p}) = \sum_{k \ge 0}^{} \Lambda^k f_{n,k}(\mathbf{p}).
\end{equation}
This statement can be easily proved by induction. It is evident that this assertion is valid for the base case \eqref{eq:96}, as well as for the subsequent iteration \eqref{eq:101}. It remains to consider an induction step. For the sake of argument, let us suppose that \eqref{eq:104} in fact holds for some \(n\). If this assertion is valid then for \(n + 1\) we have
\begin{equation}
\label{eq:102}
\Lambda \frac{\partial }{\partial \Lambda} \Lambda \sum_{k\ge 0}^{} \Lambda^k f_{n,k}(\mathbf{p})=
\Lambda \sum_{k\ge 0}^{} \left[ \left( \frac{\partial }{\partial \Lambda} \Lambda^{k+1} \right) f_{n,k}(\mathbf{p}) +
\left( \frac{\partial }{\partial \Lambda} f_{n,k}(\mathbf{p})  \right) \Lambda^{k+1}\right]. 
\end{equation}
Again, we know the matrix calculus rules for computing all the derivatives here
\begin{dmath}
\label{eq:103}
\Lambda \frac{\partial }{\partial \Lambda}  \Lambda \sum_{k \ge 0}^{}\Lambda^k f_{n,k}(\mathbf{p}) =
\sum_{k \ge 0}^{} \left( \sum_{m = 0}^{k}\Lambda^{k - m + 1} p_m f_{n,k}(\mathbf{p}) +
\sum_{m \ge 1}^{} m \Lambda^{m + k + 1} \frac{\partial }{\partial p_m} f_{n,k}(\mathbf{p}) \right) =
\sum_{k\ge 1}^{}\Lambda^k \left( \sum_{m \ge 0}^{}p_{m} f_{n,k + m - 1}(\mathbf{p}) + \sum_{m = 1}^{k-1}m \frac{\partial }{\partial p_m} f_{n,k-m-1}(\mathbf{p})\right) = \sum_{k \ge 0}^{} \Lambda^k f_{n+1,k}(\mathbf{p}).
\end{dmath}
The expected structure was obtained. Q.E.D. What can be inferred from this proposition? In fact, great deal can be inferred. From the relation
\begin{equation}
\label{eq:105}
\left( \Lambda \frac{\partial }{\partial \Lambda} \Lambda \right)^n f(\mathbf{p}) = \sum_{k}^{}\Lambda^k \widetilde{W}_{k-n}^{(n)} f(\mathbf{p}) = \sum_{k \ge 0}^{} \Lambda^k f_{n, k} (\mathbf{p}) 
\end{equation}
it can be concluded that for a non-negative integer \(n\) and \(f(\mathbf{p})\), all the series coefficients in front of \(\Lambda^k\) for \(k < 0\) should vanish. In other words,
\begin{equation}
\label{eq:106}
\widetilde{W}_k^{(n)} = 0 \qquad \text{for} \qquad k + n < 0.  
\end{equation}
With this in mind, we can now transform the non-recursive definition into a more analyzable form, namely,
\begin{equation}
\label{eq:107}
\left( \Lambda \frac{\partial }{\partial \Lambda} \Lambda \right)^n f(\mathbf{p}) = \sum_{k \ge 0}^{}\Lambda^k \widetilde{W}_{k-n}^{(n)} f(\mathbf{p}). 
\end{equation}
It should be noted that the base of the recursive definition was already obtained in \eqref{eq:97}, and the remaining task is to identify the step of the recursion. For the sake of argument, let us assume that the \(\widetilde{W}_k^{(n)}\)-s are known. To find \(\widetilde{W}_k^{(n + 1)}\) we must consider the expression
\begin{equation}
\label{eq:98}
\left( \Lambda \frac{\partial }{\partial \Lambda} \Lambda \right)^{n + 1} f(\mathbf{p}) = 
\sum_{k \ge 0}^{} \Lambda^k \widetilde{W}_{k-n-1}^{(n + 1)} f(\mathbf{p}). 
\end{equation}
On the other hand
\begin{equation}
\label{eq:108}
\left( \Lambda \frac{\partial }{\partial \Lambda} \Lambda \right)^{n + 1} f(\mathbf{p}) = \left( \Lambda \frac{\partial }{\partial \Lambda} \Lambda \right) \left( \Lambda \frac{\partial }{\partial \Lambda} \Lambda \right)^n f(\mathbf{p}) =
\Lambda \frac{\partial }{\partial \Lambda} \Lambda \sum_{k\ge 0}^{} \Lambda^k \widetilde{W}_{k-n}^{(n)} f(\mathbf{p}) .
\end{equation}
By using the expression \eqref{eq:103} for \(f_{n,k}(\mathbf{p})\) which are equal to \(\widetilde{W}_{k-n}^{(n)} f(\mathbf{p})\), we can obtain
\begin{equation}
\label{eq:109}
\left( \Lambda \frac{\partial }{\partial \Lambda} \Lambda \right)^{n+1} f(\mathbf{p})=\sum_{k \ge 1}^{}\Lambda^k
\left( \sum_{m\ge 0}^{} p_m \widetilde{W}^{(n)}_{k + m - n -1} + \sum_{m = 1}^{k - 1} m \frac{\partial }{\partial p_m} \widetilde{W}^{(n)}_{k - m - n - 1}\right) f(\mathbf{p}).
\end{equation}
Therefore, we have
\begin{equation}
\label{eq:110}
\sum_{k \ge 0}^{} \Lambda^k \widetilde{W}_{k-n-1}^{(n + 1)} f(\mathbf{p}) = \sum_{k \ge 1}^{}\Lambda^k
\left( \sum_{m\ge 0}^{} p_m \widetilde{W}^{(n)}_{k + m - n -1} + \sum_{m = 1}^{k - 1} m \frac{\partial }{\partial p_m} \widetilde{W}^{(n)}_{k - m - n - 1}\right) f(\mathbf{p}).
\end{equation}
The first immediate consequence is that the coefficient in front of \(\Lambda^0\) on the left-hand side vanishes, i.e.
\begin{equation}
\label{eq:111}
\widetilde{W}^{(n + 1)}_{k} = 0 \qquad \text{for} \qquad k + n + 1 = 0. 
\end{equation}
It is noteworthy that the upper index of \(\widetilde{W}\) in this relation commences from 1, because \(\widetilde{W}_0^{(0)}\ne 0\) (see eq. \eqref{eq:97}). A second consequence arises from a term-by-term comparison of the remaining coefficients preceding \(\Lambda^k\)
\begin{equation}
\label{eq:112}
\widetilde{W}_k^{(n + 1)} = \sum_{m \ge 0}^{} p_m \widetilde{W}_{k + m}^{(n)} +
\sum_{m = 1}^{k + n} m \frac{\partial }{\partial p_m} \widetilde{W}_{k - m}^{(n)} \qquad \text{for} \qquad k + n \ge 0. 
\end{equation}
It is the last building block in the construction of the recursive definition introduced in eqs. \eqref{eq:42} to \eqref{eq:48}.
\subsection{\(m = -1\) integer ray}
\label{sec:org39787ee}
Another family of the \(\widetilde{W}\) algebras first appeared in \cite{mironov-1996-unitar-matrix}
in the context of the generalized Kontsevich model. One phase of this model corresponds to the described \(\widetilde{W}^{(n)}\) algebras, but the other differs. It is therefore appropriate to switch the notation slightly and rename \(\widetilde{W}^{(n)}\) to \(\widetilde{W}^{(+ , n)}\). The newly discovered algebra has been given the name \(\widetilde{W}^{(-, n)}\). It should be noted that the sign convention differs from that used in the original paper. The corresponding partition functions, \(Z_{-n}(\mathbf{p}, \mathbf{g})\), \(n \ge 1\), were rewritten in a more concise form in \cite{mironov-2023-inter-matrix,mironov-2023-kp-integ}
\begin{equation}
\label{eq:115}
Z_{-n} = \iint dXdY \exp \tr \left(\frac{Y^n}{n} + Y\Lambda - YX + \sum_{k} \frac{g_kX^{k}}{k}
\right) .
\end{equation}
The \(W\)-representation of these models reads as
\begin{equation}
\label{eq:155}
Z_{-n} = e^{\frac{1}{n}H_{-n}^{(1)}} \cdot e^{\sum_{k}^{} \frac{g_k p_k}{k}}.
\end{equation}
The question of finding the Ward identities is much more subtle here.
For an \(n = 2\) case, the Gaussian integral in eq. \eqref{eq:115} over the matrix \(Y\) can be taken, resulting in
\begin{equation}
\label{eq:130}
Z_2 = e^{- \frac{1}{2} p_2} \int dX\, \exp \tr \left( - \frac{X^2}{2} + X\Lambda + \sum_{k}^{} \frac{g_k X^k}{k} \right).
\end{equation}
This is precisely the generalized Kontsevich model in the character phase. It satisfies the Ward identities
\begin{equation}
\label{eq:132}
\left( g_1 \delta_{n,1} + \delta_{n,2} - n \frac{\partial }{\partial p_n} + \sum_{k > 1}^{} g_k \widetilde{W}_{k + n - 2}^{(-, k - 1)} \right) e^{\frac{p_2}{2}} Z_2 = 0, \qquad n\ge1.
\end{equation}
The \(\widetilde{W}^{(-, n)}_k\) operators used here can be non-recursively defined as follows:
\begin{equation}
\label{eq:117}
\left( \frac{\partial }{\partial \Lambda}\right)^{n} f(\mathbf{p})= \sum_{k}^{}\Lambda^{k} \widetilde{W}_{k+n}^{(-, n)} f (\mathbf{p}).
\end{equation}
The corresponding recursive definition is also known
\begin{equation}
\label{eq:126}
  \widetilde{W}_k^{(-,n + 1)} =   \sum_{m \ge 0}^{} p_m \widetilde{W}_{k + m}^{(-,n)} + \sum_{m = 1}^{k - n} m \frac{\partial }{\partial p_m}
  \widetilde{W}_{k - m}^{(-,n)} \qquad \text{for} \qquad k \ge n + 1,  
\end{equation}
\begin{equation}
\label{eq:127}
\widetilde{W}_k^{(-,n + 1)} = 0, \qquad \text{otherwise},
\end{equation}
with the base
\begin{equation}
\label{eq:128}
\widetilde{W}_k^{(-, 0)}=\delta_{k,0}.
\end{equation}
This result can be obtained through a pure analogy with the case of the \(\widetilde{W}^{(+,n)}\) algebras. The aforementioned reasoning allows us to set the lower bound for the sum
\begin{equation}
\label{eq:129}
\left( \frac{\partial }{\partial \Lambda}  \right)^n f(\mathbf{p}) = \sum_{k \ge 0}^{}\Lambda^k \widetilde{W}_{k + n}^{(-,n)} f(\mathbf{p}).
\end{equation}
This implies eq. \eqref{eq:127}.
The base of the recursion (eq. \eqref{eq:128}) again can be calculated from
\begin{equation}
\label{eq:131}
\left( \frac{\partial }{\partial \Lambda}  \right)^0 f(\mathbf{p}) = I f(\mathbf{p})= \sum_{k \ge 0}^{}
\Lambda^k \widetilde{W}_k^{(-, 0)} f (\mathbf{p}).
\end{equation}
To evaluate the step of the recursion, one should consider the following chain of equalities, analogous to the \(\widetilde{W}^{(+, n)}\) case, 
\begin{dmath}
\label{eq:133}
\sum_{k \ge 0}^{} \Lambda^k \widetilde{W}_{k + n + 1}^{(-, n + 1)} f(\mathbf{p}) =
\left( \frac{\partial }{\partial \Lambda}  \right)^{n + 1} f(\mathbf{p}) = \left( \frac{\partial }{\partial \Lambda}  \right)
\left( \frac{\partial }{\partial \Lambda}  \right)^n f(\mathbf{p}) =
\frac{\partial }{\partial \Lambda} \sum_{k \ge 0}^{}\Lambda^k \widetilde{W}_{k + n}^{(-, n)} f(\mathbf{p}) =
\sum_{k \ge 0}^{} \left( \sum_{m \ge 0}^{k - 1}\Lambda^{k - m - 1}p_m + \sum_{m \ge 1}^{}m \Lambda^{k + m - 1} \frac{\partial }{\partial p_m} \right)\widetilde{W}_{k + n}^{(-, n)} f(\mathbf{p}) =
\sum_{k \ge 0}^{} \Lambda^k \left( \sum_{m \ge 0}^{} p_m \widetilde{W}_{k + m + n + 1}^{(-, n)} +
\sum_{m = 1}^{k + 1} m \frac{\partial }{\partial p_m} \widetilde{W}_{k - m + n + 1}^{(-, n)} \right) f(\mathbf{p})
\end{dmath}
A term-by-term comparison of the first and the last expressions in this chain yields eq. \eqref{eq:126}. Next, some examples are in order. The initial step of the recursion for \(k \ge 1\) is as follows 
\begin{dmath}
\label{eq:135}
\widetilde{W}_k^{(-, 1)} = \sum_{m \ge 0}^{} p_m \delta_{k + m, 0} + \sum_{m = 1}^{k} m \frac{\partial }{\partial p_m} \delta_{k - m, 0} = k \frac{\partial }{\partial p_k}.
\end{dmath}
The absence of the \(k = 0\) term in comparison to the \(\widetilde{W}_k^{(+, 1)}\) case (eq. \eqref{eq:124}) is solely due to the difference between the vanishing conditions (eqs. \eqref{eq:121} and \eqref{eq:127}).
Consequently, for \(k \ge 2\), we have the following:
\begin{equation}
\label{eq:136}
\widetilde{W}_k^{(-, 2)} = \sum_{m \ge 0}^{} (k + m)p_m  \frac{\partial }{\partial p_{k + m}}
+ \sum_{m = 1}^{k - 1} m (k - m)\frac{\partial^2 }{\partial p_m \partial p_{k - m}}.
\end{equation}
\subsection{\(n = 1\) algebras}
\label{sec:org5ac6e7d}
In \cite{alexandrov-2014-kp-integ}, another family of the \(\widetilde{W}\) algebras was introduced. The following partition functions \(Z^{(m)}= Z^{(m)}(\mathbf{p})\) were discussed here:
\begin{equation}
\label{eq:156}
Z^{(m)}= \int \exp \tr \left( Y_m + \sum_{k}^{} \frac{p_k  X_1^k}{k} \right) \frac{\prod_{i = 1}^{m}  e^{- \tr X_i Y_i}dX_i dY_i
}{\prod_{i = 1}^{m - 1} \det \left( I \otimes I - Y_{i} \otimes X_{i + 1} \right)
}, 
\end{equation}
which possess the \(W\)-representation
\begin{equation}
\label{eq:158}
Z^{(m)} = e^{H_1^{(m)}}\cdot 1,
\end{equation}
and satisfy the Ward identities
\begin{equation}
\label{eq:157}
\widetilde{W}_k^{(m, +)} Z^{(m)} = (k + 1) \frac{\partial Z^{(m)}}{\partial p_{k + 1}}, \qquad k\ge 0.
\end{equation}
The non-recursive definition for the \(\widetilde{W}_{k}^{(m, +)}\)-operators reads as
\begin{equation}
\label{eq:134}
 \left( \Lambda \frac{\partial }{\partial \Lambda} \right)^m \Lambda f(\mathbf{p}) = \sum_{k}^{} \Lambda^k \widetilde{W}_{k-1}^{(m, +)}f(\mathbf{p}).
\end{equation}
It is opportune to recall what one can obtain after taking a trace of such an expression
\begin{equation}
\label{eq:137}
H_1^{(m)} = \tr \left( \left( \Lambda \frac{\partial }{\partial \Lambda}  \right)^m\Lambda \right)
= \sum_{k}^{}p_k \widetilde{W}_{k - 1}^{(m, + )}.
\end{equation}
But, on the other hand
\begin{equation}
\label{eq:138}
\left( \Lambda \frac{\partial }{\partial \Lambda}  \right)^{m }\Lambda = \Lambda \left( \frac{\partial }{\partial \Lambda} \Lambda \right)^m. 
\end{equation}
Thus,
\begin{equation}
\label{eq:139}
\Lambda \left( \frac{\partial }{\partial \Lambda} \Lambda \right)^m f(\mathbf{p}) = \sum_{k}^{}\Lambda^k \widetilde{W}_{k - 1}^{(m, +)} f(\mathbf{p}).
\end{equation}
Moreover,
\begin{dmath}
\label{eq:140}
\left( \frac{\partial }{\partial \Lambda} \Lambda \right)^m f(\mathbf{p}) = \sum_{k}^{}\Lambda^{k - 1} \widetilde{W}_{k - 1}^{(m, +)} f(\mathbf{p}) = \sum_{k}^{}\Lambda^k \widetilde{W}_k^{(m, +)} f(\mathbf{p}).
\end{dmath}
The base of the recursive definition, as the immediate consequence of the aforementioned relation, is
\begin{equation}
\label{eq:141}
\widetilde{W}_k^{(m, +)} = \delta_{k, 0}.
\end{equation}
It is also evident that \(\widetilde{W}_k^{(m, +)}\) vanishes for \(k < 0\). The chain of equalities to find the recursion step for this case is as follows:
\begin{dmath}
\label{eq:142}
\sum_{k \ge 0}^{} \Lambda^k \widetilde{W}_{k}^{(m + 1, +)} f(\mathbf{p}) =
\left( \frac{\partial }{\partial \Lambda}  \Lambda \right)^{m + 1} f(\mathbf{p}) = \left( \frac{\partial }{\partial \Lambda}  \Lambda\right)
\left( \frac{\partial }{\partial \Lambda}  \Lambda \right)^m f(\mathbf{p}) =
\frac{\partial }{\partial \Lambda}\Lambda \sum_{k \ge 0}^{}\Lambda^k \widetilde{W}_{k}^{(m, +)} f(\mathbf{p}) =
\sum_{k \ge 0}^{} \left( \sum_{n \ge 0}^{k}\Lambda^{k - n}p_n + \sum_{n \ge 1}^{}n \Lambda^{k + n} \frac{\partial }{\partial p_n} \right)\widetilde{W}_{k}^{(m, +)} f(\mathbf{p}) =
\sum_{k \ge 0}^{} \Lambda^k \left( \sum_{n \ge 0}^{} p_n \widetilde{W}_{k + n}^{(m,+)} +
\sum_{n = 1}^{k} n \frac{\partial }{\partial p_n} \widetilde{W}_{k - n}^{(m, +)} \right) f(\mathbf{p}).
\end{dmath}
Therefore, the recursion step for the \(\widetilde{W}^{(m, +)}_k\) operators for \(k \ge 0\) is
\begin{equation}
\label{eq:143}
\widetilde{W}_k^{(m + 1, +)} = \sum_{n \ge 0}^{}p_n \widetilde{W}_{k + n}^{(m, +)}
+ \sum_{n = 1}^{k} n \frac{\partial }{\partial p_{n}} \widetilde{W}_{k - n}^{(m, +)}.
\end{equation}
For \(k \ge 0\) we have
\begin{equation}
\label{eq:145}
\widetilde{W}_k^{(1, +)} = N \delta_{k,0} + k \frac{\partial }{\partial p_k},
\end{equation}
and
\begin{dmath}
\label{eq:146}
\widetilde{W}_k^{(2, +)} = N^2 \delta_{k,0} + \sum_{m \ge 0}^{}(k + m)p_m \frac{\partial }{\partial p_{k + m}} + Nk \frac{\partial }{\partial p_k} + \sum_{m = 1}^{k - 1} m (k - m) \frac{\partial ^2}{\partial p_m \partial p_{k - m}}.  
\end{dmath}
\subsection{\(n = -1\) algebras}
\label{sec:orgbe26feb}
The final piece of the introductory material is not, in fact, about the known \(\widetilde{W}\) algebras. Instead, it concerns the simple counterparts of the \(\widetilde{W}^{(m, +)}\) algebras, which are essential to consider before moving on to much more sophisticated examples. They are defined as
\begin{equation}
\label{eq:147}
\Lambda^{-1}\left( \Lambda \frac{\partial }{\partial \Lambda}  \right) ^m f(\mathbf{p})= \sum_{k}^{}\Lambda^k \widetilde{W}_{k + 1}^{(m, -)} f(\mathbf{p}),
\end{equation}
and, in terms of the generalized picture, correspond to the following operators:
\begin{equation}
\label{eq:148}
H_{-1}^{(-m)} = \tr \left( \Lambda^{-1} \left( \Lambda \frac{\partial }{\partial \Lambda}  \right)^m \right).
\end{equation}
Their definition can be rewritten in a more convenient form
\begin{dmath}
\label{eq:149}
\left( \Lambda \frac{\partial }{\partial \Lambda}  \right)^m f(\mathbf{p}) = \sum_{k}^{}\Lambda^{k + 1} \widetilde{W}_{k + 1}^{(m, -)} f(\mathbf{p}) = \sum_{k}^{}\Lambda^k \widetilde{W}_k^{(m, -)} f(\mathbf{p}).
\end{dmath}
By considering the case \(m = 0\), one can derive the base of the recursion
\begin{equation}
\label{eq:144}
\widetilde{W}_k^{(0, -)} = \delta_{k,0}.
\end{equation}
The negative powers of \(\Lambda\) cannot appear on the left-hand side of \eqref{eq:149}. Consequently, we have the expansion
\begin{equation}
\label{eq:151}
\left( \Lambda \frac{\partial }{\partial \Lambda}  \right)^m f(\mathbf{p}) = \sum_{k \ge 0}^{} \Lambda^k\widetilde{W}_k^{(m,-)}
f(\mathbf{p}).
\end{equation}
Our cherished chain of equalities
\begin{dmath}
\label{eq:152}
\sum_{k \ge 0}^{} \Lambda^k \widetilde{W}_{k}^{(m + 1, -)} f(\mathbf{p}) =
\left( \Lambda\frac{\partial }{\partial \Lambda}   \right)^{m + 1} f(\mathbf{p}) = \left( \Lambda\frac{\partial }{\partial \Lambda}  \right)
\left( \Lambda\frac{\partial }{\partial \Lambda}   \right)^m f(\mathbf{p}) =
\Lambda\frac{\partial }{\partial \Lambda} \sum_{k \ge 0}^{}\Lambda^k \widetilde{W}_{k}^{(m, -)} f(\mathbf{p}) =
\sum_{k \ge 0}^{} \left( \sum_{n \ge 0}^{k - 1}\Lambda^{k - n}p_n + \sum_{n \ge 1}^{}n \Lambda^{k + n} \frac{\partial }{\partial p_n} \right)\widetilde{W}_{k}^{(m, -)} f(\mathbf{p}) =
\sum_{k \ge 1}^{} \Lambda^k \left( \sum_{n \ge 0}^{} p_n \widetilde{W}_{k + n}^{(m,-)} +
\sum_{n = 1}^{k} n \frac{\partial }{\partial p_n} \widetilde{W}_{k - n}^{(m, -)} \right) f(\mathbf{p})
\end{dmath}
demonstrates that for \(k \ge 1\) we have
\begin{equation}
\label{eq:153}
\widetilde{W}_k^{(m + 1, -)} = \sum_{n \ge 0}^{} p_n \widetilde{W}_{k + n}^{(m, -)} +
\sum_{n = 1}^{k }n \frac{\partial }{\partial p_n}  \widetilde{W}_{k - n}^{(m, -)}
\end{equation}
and otherwise
\begin{equation}
\label{eq:154}
\widetilde{W}_k^{(m + 1, -)} = 0.
\end{equation}
By employing the aforementioned recursion procedure, one can obtain for \(k \ge 1\)
\begin{dmath}
\label{eq:155}
\widetilde{W}_k^{(1, -)} =  k \frac{\partial }{\partial p_k}
\end{dmath}
and
\begin{equation}
\label{eq:156}
\widetilde{W}_k^{(2, -)} = \sum_{m \ge 0}^{} (k + m)p_m  \frac{\partial }{\partial p_{k + m}}
+ \sum_{m = 1}^{k - 1} m (k - m)\frac{\partial^2 }{\partial p_m \partial p_{k - m}}.
\end{equation}
At first glance these operators don't differ from eqs. \eqref{eq:135} and \eqref{eq:136}, but the discrepancy is obscured by the restrictions placed on \(k\). For example, the quantity \(\widetilde{W}_k^{(2, -)}\) is defined to be non-zero for \(k = 1\), but this is not the case for the \(\widetilde{W}_k^{(-, 2)}\).
\section{Generalized \(\widetilde{W}\) algebras}
\label{sec:orgbac4350}
Recently, the WLZZ models \cite{wang-2022-super-partit} and their generalization \cite{mironov-2023-kp-integ} were discovered, bringing to the stage the \(W\)-representation of the \(Z_k^{(m)}\) partition functions, namely
\begin{equation}
\label{eq:19}
Z_n^{(m)} = \begin{cases}
  e^{\frac{1}{n} H_{n}^{(m)}} \cdot 1, & n \ge 1,\\
e^{-\frac{1}{n} H_n^{(m)}} \cdot e^{\sum_k \frac{g_k p_k}{k}}, & n \le -1.
\end{cases}
\end{equation}
where \(n\) and \(m\) are always of the same sign. These models represent a comprehensive generalization of the examples presented in the section \ref{sec:org850b9ed} with \(Z_n^{(1)} = Z_n\), \(Z_{-n}^{(-1)} = Z_{-n}\) and \(Z^{(m)}_1 \equiv Z^{(m)}\). Half of them already possess matrix integral representation
\begin{equation}
Z^{(m)}_n= \int \exp \tr \left( Y_m^n + \sum_{k}^{} \frac{p_k  X_1^k}{k} \right) \frac{\prod_{i = 1}^{m}  e^{- \tr X_i Y_i}dX_i dY_i
}{\prod_{i = 1}^{m - 1} \det \left( I \otimes I - Y_{i} \otimes X_{i + 1} \right)
},   
\end{equation}
for both \(m\) and \(n\) positive. As previously stated in section \ref{sec:org83d5c97}, the objective is to find the \(\widetilde{W}^{(\pm m,\pm n)}_k\) operators that satisfy the following identities:
\begin{equation}
\label{eq:24}
H_{-n}^{(-m)} = \tr \left(\Lambda^{-1} \left( \Lambda \frac{\partial }{\partial \Lambda}  \right)^m  \right)^{n} =
\sum_{k }^{} p_k \widetilde{W}_{k + n}^{(-m, -n)},
\end{equation}
\begin{equation}
\label{eq:150}
H_n^{(m)}= \tr \left( \left( \Lambda \frac{\partial }{\partial \Lambda}  \right)^m \Lambda \right)^{n} = \sum_{k}^{}p_k \widetilde{W}_{k-n}^{(m,n)}.
\end{equation}
Based on the information presented in section \ref{sec:org850b9ed}, it can be conjectured that these relations hold not only in traced form, but the following is also true
\begin{equation}
\label{eq:159}
\left(\Lambda^{-1} \left( \Lambda \frac{\partial }{\partial \Lambda}  \right)^m  \right)^{n} f(\mathbf{p})=
\sum_{k }^{} \Lambda^{k} \widetilde{W}_{k + n}^{(-m, -n)} f(\mathbf{p}), 
\end{equation}
\begin{equation}
\label{eq:160}
\left( \left( \Lambda \frac{\partial }{\partial \Lambda}  \right)^m \Lambda \right)^{n} f(\mathbf{p})= \sum_{k}^{}\Lambda^k\widetilde{W}_{k-n}^{(m,n)} f (\mathbf{p}). 
\end{equation}
Starting from these definitions, one can derive recursive ones, as demonstrated in subsection \ref{sec:orgec1c787} for an \(n \ge 1\) case and in subsection \ref{sec:org8432816} for an \(n \le -1\) case respectively.
\subsection{\(n \ge 1\) algebras}
\label{sec:orgec1c787}
\begin{figure}[h]
\centering
\begin{tikzpicture}
     \tikzset{
         every pin/.style={fill=gray!30!white,rectangle,rounded corners=3pt,font=\tiny},
         small dot/.style={fill=black,circle,scale=0.3},
     }
\draw[->, thick] (1,1) -- (2,1) node[above] {};
 \draw[->, thick] (2,1) -- (3,1) node[above] {};
 \node at (3.5,1) {\ldots};
 \draw[->, thick] (4,1) -- (5,1) node[above] {};
 \draw[->, thick] (5,1) -- (1,2) node[above] {};
 \draw[->, thick] (1,2) -- (2,2) node[above] {};
 \draw[->, thick] (2,2) -- (3,2) node[above] {};
 \node at (3.5,2) {\ldots}; 
 \draw[->, thick] (4,2) -- (5,2) node[above] {};

 \draw[->, thick] (5,2) -- (1,3) node[above] {};
 \draw[->, thick] (1,3) -- (2,3) node[above] {};
 \draw[->, thick] (2,3) -- (3,3) node[above] {};
 \node at (3.5,3) {\ldots};
 \draw[->, thick] (4,3) -- (5,3) node[above] {};
 \node at (3,4) {\(\vdots\)};
 \draw[->, thick] (5,3) -- (1,4) node[above] {};
 \draw[->, thick] (5,4) -- (1,5) node[above] {};
 \node [pin=150:{\( \widetilde{W}_{k}^{(m,0)} = \widetilde{W}_k^{(m,1\vert0)} \)}]     at (1,1)      {};
 \node [pin=30:{\( \widetilde{W}_k^{(m,1\vert m)} \)}]      at (5,1)      {};
 \node [pin=30:{\( \widetilde{W}_k^{(m,2\vert m)} \)}]      at (5,2)      {};
 \node [pin=30:{\( \widetilde{W}_k^{(m,3\vert m)} \)}]      at (5,3)      {};
 \node [pin=30:{\( \widetilde{W}_k^{(m,n\vert m)} \)}]      at (5,4)      {};
 \node [pin=150:{\(\widetilde{W}_{k}^{(m,1)} = \widetilde{W}_k^{(m,2\vert0)}\) }]      at (1,2)      {};
 \node [pin=30:{\( \widetilde{W}_{k}^{(m,n)} = \widetilde{W}_k^{(m,n + 1\vert0)}\) }]      at (1,5)      {};
 \node [pin=150:{\( \widetilde{W}_{k}^{(m,2)} = \widetilde{W}_k^{(m,3\vert0)}\) }]      at (1,3)      {};
 \node [pin=150:{\( \widetilde{W}_{k}^{(m,3)} = \widetilde{W}_k^{(m,4\vert0)}\) }]      at (1,4)      {};
 \node [pin=-120:{\( \widetilde{W}_k^{(m,1\vert1)}\) }]      at (2,1)      {};
 \node [pin=-90:{\( \widetilde{W}_k^{(m,1\vert2)}\) }]      at (3,1)      {};
 \node [pin=-60:{\( \widetilde{W}_k^{(m,1\vert m - 1)}\) }]      at (4,1)      {};
\end{tikzpicture}
\caption[]{Structure of recursive procedure for \( \widetilde{W}_k^{(m,n)} \) operators}
\label{fig:2}
\end{figure}
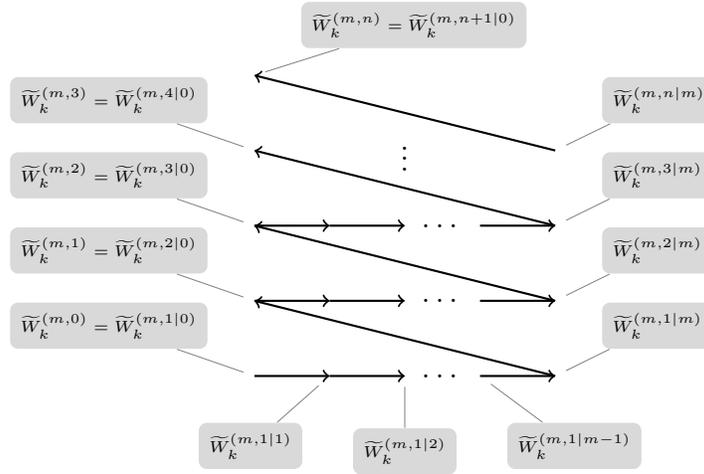

The starting point is the definition
\begin{equation}
\label{eq:161}
\left( \left( \Lambda \frac{\partial }{\partial \Lambda}  \right)^m \Lambda \right)^{n} f(\mathbf{p})= \sum_{k}^{}\Lambda^k\widetilde{W}_{k-n}^{(m,n)} f(\mathbf{p}).
\end{equation}
The novel aspect of this expression is that two distinct repeated operations must be considered in order to construct a recursion: applying operator \(\frac{\partial }{\partial \Lambda} \Lambda\) and regular multiplication by \(\Lambda\). To address this issue we propose the following auxiliary operators \(\widetilde{W}_k^{(m, n | l)}\), defined as:
\begin{equation}
\label{eq:164}
\Lambda \left( \frac{\partial }{\partial \Lambda} \Lambda \right)^l \left( \Lambda \left( \frac{\partial }{\partial \Lambda} \Lambda \right)^m \right)^{n - 1} f(\mathbf{p}) = \sum_{k}^{} \Lambda^k \widetilde{W}_{k - n}^{(m, n| l)} f(\mathbf{p}), 
\end{equation}
what means that
\begin{dmath}
\label{eq:173}
\left( \frac{\partial }{\partial \Lambda} \Lambda \right)^l \left( \Lambda \left( \frac{\partial }{\partial \Lambda} \Lambda \right)^m \right)^{n - 1} f(\mathbf{p}) = \sum_{k}^{} \Lambda^{k - 1} \widetilde{W}_{k - n}^{(m, n| l)} f(\mathbf{p}) = \sum_{k}^{} \Lambda^{k} \widetilde{W}_{k - n + 1}^{(m, n| l)} f(\mathbf{p}).
\end{dmath}
Our desired \(\widetilde{W}_k^{(m,n)}\)-s are, in fact, merely the special case of \(\widetilde{W}_k^{(m, n | l)}\) operators
\begin{equation}
\label{eq:181}
\widetilde{W}_k^{(m,n)} = \widetilde{W}_k^{(m, n + 1 | 0)}.
\end{equation}
As demonstrated in the previous section, by examining the left-hand side of such a relation, it is evident that only the non-negative powers of \(\Lambda\) can appear in the expansion on the right-hand side, i.e.
\begin{equation}
\label{eq:179}
\widetilde{W}_k^{(m,n | l)} = 0 \qquad \text{for} \qquad n + k \le 0, 
\end{equation}
and
\begin{equation}
\label{eq:180}
\left( \frac{\partial }{\partial \Lambda} \Lambda \right)^l \left( \Lambda \left( \frac{\partial }{\partial \Lambda} \Lambda \right)^m \right)^{n - 1} f(\mathbf{p}) = \sum_{k \ge 0}^{} \Lambda^{k} \widetilde{W}_{k - n + 1}^{(m, n| l)} f(\mathbf{p}).
\end{equation}
By considering the case of \(l = 0\) and \(n = 1\) one can derive the base of the recursion
\begin{equation}
\label{eq:178}
\widetilde{W}_k^{(m, 1 | 0)} = \delta_{k, 0}.
\end{equation}
The boundary cases \(l = m\) and \(l = 0\) are of particular interest here. For the former case we have
\begin{dmath}
\label{eq:167}
\left( \Lambda \left( \frac{\partial }{\partial \Lambda} \Lambda \right)^m \right)^{n} f(\mathbf{p}) =
\sum_{k \ge 1}^{} \Lambda^k \widetilde{W}_{k - n}^{(m, n| m)} f(\mathbf{p}),
\end{dmath}
while for the latter
\begin{dmath}
\label{eq:166}
\left( \Lambda \left( \frac{\partial }{\partial \Lambda} \Lambda \right)^m \right)^{n} f(\mathbf{p}) = \sum_{k \ge0}^{} \Lambda^{k} \widetilde{W}_{k - n}^{(m, n + 1| 0)} f(\mathbf{p}).
\end{dmath}
The comparison of eq. \eqref{eq:167} and \eqref{eq:166} allows us to conclude that
\begin{equation}
\label{eq:183}
\widetilde{W}_{k}^{(m, n + 1| 0)} = \widetilde{W}_{k}^{(m, n | m)}  \qquad \text{for} \qquad
n + k > 0,
\end{equation}
\begin{equation}
\label{eq:184}
\widetilde{W}_{k}^{(m, n + 1| 0)} = 0 \qquad \text{otherwise.}
\end{equation}
It is now evident what effect multiplying by \(\Lambda\) has. The final step is to address the issue of the \(\frac{\partial }{\partial \Lambda} \Lambda\) repeating action. As is customary, the following chain of equalities is employed to facilitate the solution:
\begin{dmath}
\label{eq:171}
\sum_{k \ge 0}^{} \Lambda^{k} \widetilde{W}_{k - n + 1}^{(m, n\mid l + 1)}  f(\mathbf{p}) =
\left( \frac{\partial }{\partial \Lambda} \Lambda \right)^{l + 1 } \left( \Lambda \left( \frac{\partial }{\partial \Lambda} \Lambda \right)^m \right)^{n - 1} f(\mathbf{p}) = \left( \frac{\partial }{\partial \Lambda} \Lambda \right) \left( \frac{\partial }{\partial \Lambda} \Lambda \right)^{l } \left( \Lambda \left( \frac{\partial }{\partial \Lambda} \Lambda \right)^m \right)^{n - 1} f(\mathbf{p}) =
\frac{\partial }{\partial \Lambda}\Lambda \sum_{k \ge 0}^{}\Lambda^{k} \widetilde{W}_{k - n + 1}^{(m, n | l)} f(\mathbf{p}) =
\sum_{k \ge 0}^{} \left( \sum_{r \ge 0}^{k}\Lambda^{k - r}p_r + \sum_{r \ge 1}^{}r \Lambda^{k + r} \frac{\partial }{\partial p_r} \right)\widetilde{W}_{k - n + 1}^{(m, n | l)} f(\mathbf{p}) =
\sum_{k \ge 0}^{} \Lambda^k \left( \sum_{r \ge 0}^{} p_r \widetilde{W}_{k - n + r + 1}^{(m, n | l)} +
\sum_{r = 1}^{k} r \frac{\partial }{\partial p_r} \widetilde{W}_{k - n - r + 1}^{(m, n | l)} \right) f(\mathbf{p}).
\end{dmath}
A term-by-term comparison of the first expression with the last entails
\begin{equation}
\label{eq:182}
\widetilde{W}_{k}^{(m,n | l + 1)} = \sum_{r \ge 0}^{}p_r \widetilde{W}_{k + r}^{(m,n |l)}
+ \sum_{r = 1}^{k + n - 1} r \frac{\partial }{\partial p_r}  \widetilde{W}_{k - r}^{(m, n |l)} \qquad \text{for}\qquad n + k > 0.
\end{equation}
This is the final piece of the recursive procedure that was desired. To clarify,
we will summarize the results in a closed form
\begin{subequations}
\label{eq:185}
\begin{empheq}[box=\fbox]{gather}
\label{eq:187}
\widetilde{W}_k^{(m,n)} = \widetilde{W}_k^{(m, n + 1 | 0)}, \\
\label{eq:188}
  \widetilde{W}_{k}^{(m,n | l + 1)} = \sum_{r \ge 0}^{}p_r \widetilde{W}_{k + r}^{(m,n |l)}
+ \sum_{r = 1}^{k + n - 1} r \frac{\partial }{\partial p_r}  \widetilde{W}_{k - r}^{(m, n |l)} \qquad \text{for}\qquad n + k > 0, \\
\label{eq:186}
  \widetilde{W}_{k}^{(m, n + 1| 0)} = \widetilde{W}_{k}^{(m, n | m)} \qquad \text{for} \qquad n > 0,\\
\widetilde{W}_k^{(m, 1 | 0)} = \delta_{k, 0}, \\
\widetilde{W}_k^{(m,n | l)} = 0 \qquad \text{otherwise.}
\end{empheq}
\end{subequations}
Figure \ref{fig:2} provides a graphical representation of the recursive procedure.

It is important to verify whether indeed \(\widetilde{W}^{(m,1)}_k = \widetilde{W}^{(m, + )}_k\) and \(\widetilde{W}_k^{(1, n)} = \widetilde{W}_k^{(+, n)}\). We will begin by calculating \(\widetilde{W}_k^{(m, 1)}\). By virtue of eq. \eqref{eq:187} we can ascertain that \(\widetilde{W}_k^{(m, 1)} = \widetilde{W}_k^{(m, 2 | 0)}\), while eq. \eqref{eq:186} entails that \(\widetilde{W}_k^{(m, 2 | 0)}\) is equal to \(\widetilde{W}_k^{(m, 1 | m)}\) for \(k \ge 0\). To obtain \(\widetilde{W}_k^{(m, 1 | m)}\), one should follow the recursion relation of eq. \eqref{eq:188}
\begin{equation}
\label{eq:189}
 \widetilde{W}_{k}^{(m,1 | l + 1)} = \sum_{r \ge 0}^{}p_r \widetilde{W}_{k + r}^{(m,1 |l)}
+ \sum_{r = 1}^{k} r \frac{\partial }{\partial p_r}  \widetilde{W}_{k - r}^{(m, 1 |l)} \qquad \text{for}\qquad k \ge 0,
\end{equation}
with the base \(\widetilde{W}_k^{(m, 1 |0)} = \delta_{k, 0}\). These are precisely the steps of the recursion relation for \(\widetilde{W}_k^{(m, +)}\).

The subsequent case is that of \(\widetilde{W}_k^{(1, n)}\) algebras. Here we have \(\widetilde{W}_k^{(1,n + 1)} = \widetilde{W}_k^{(1, n + 2 | 0)}\) and in turn \(\widetilde{W}_k^{(1, n + 2 | 0)} = \widetilde{W}_k^{(1, n + 1| 1)}\) for \(n + k \ge 0\). After that, the recursion step is as follows:
\begin{equation}
\label{eq:190}
\widetilde{W}_k^{(1, n + 1| 1)} =\sum_{r \ge 0}^{} p_r \widetilde{W}_{k + r}^{(1,n + 1|0)} + \sum_{r = 1}^{k + n} r \frac{\partial }{\partial p_r}  \widetilde{W}_{k - r}^{(1, n + 1| 0)} \qquad \text{for} \qquad
n + k \ge 0.
\end{equation}
Substitution of \(\widetilde{W}_k^{(1, n + 1 | 0)} = \widetilde{W}_k^{(1, n| 1)}\) to the expression above gives
\begin{equation}
\label{eq:191}
\widetilde{W}_k^{(1, n + 1| 1)} =\sum_{r \ge 0}^{} p_r \widetilde{W}_{k + r}^{(1,n|1)} + \sum_{r = 1}^{k + n} r \frac{\partial }{\partial p_r}  \widetilde{W}_{k - r}^{(1, n| 1)} \qquad \text{for} \qquad
n + k \ge 0.
\end{equation}
This is precisely the recursion relation \eqref{eq:112} of the \(\widetilde{W}_k^{(+, n)}\) algebra. The base of this recursion is also reproduced
\begin{equation}
\label{eq:192}
\widetilde{W}_k^{(+, n)} = \widetilde{W}_k^{(1,1|0)} = \delta_{k, 0},
\end{equation}
but the last step before the base should be interpreted in the sense of \eqref{eq:190}.

After all, by taking the trace of \eqref{eq:161} it is evident that the described \(\widetilde{W}\) operators indeed satisfy the desired relation
\begin{equation}
\label{eq:162}
H_n^{(m)}= \tr \left( \left( \Lambda \frac{\partial }{\partial \Lambda}  \right)^m \Lambda \right)^{n} = \sum_{k}^{}p_k \widetilde{W}_{k-n}^{(m,n)}.
\end{equation}
\subsection{\(n \le -1\) algebras}
\label{sec:org8432816}
It is now straightforward to find a recursive definition for the \(\widetilde{W}\) operators, defined as:
\begin{equation}
\label{eq:163}
\left( \Lambda^{-1}\left( \Lambda \frac{\partial }{\partial \Lambda}  \right)^m  \right)^{n} f(\mathbf{p})= \sum_{k}^{}\Lambda^k\widetilde{W}_{k + n}^{(-m, -n)} f(\mathbf{p}). 
\end{equation}
Analogous to the previous subsection let us consider the following auxiliary operators:
\begin{equation}
\label{eq:168}
\Lambda^{-1}\left( \Lambda \frac{\partial }{\partial \Lambda}  \right)^l \left( \Lambda^{-1}\left( \Lambda \frac{\partial }{\partial \Lambda}  \right)^m  \right)^{n - 1} f(\mathbf{p})= \sum_{k}^{}\Lambda^k\widetilde{W}_{k + n}^{(-m, -n | -l)} f(\mathbf{p}),
\end{equation}
where \(l \le m\). It is important and easy to understand that, in fact, when evaluating such operators, you would never face the problem of taking a matrix derivative of the inverse matrix. Nevertheless, in the case of \(m = 0\) the only power of \(\Lambda\) that will appear on the left-hand side is \(\Lambda^{-n}\). Similarly, in the case of \(l = 0\) the \(\Lambda^{-1}\) term can be presented. The case of \(m = 0\) is self-evident, and therefore will not be considered here. For other values of \(m\), the following restriction can be formulated:
\begin{equation}
\label{eq:169}
\widetilde{W}_k^{(-m, -n| -l)} = 0 \qquad \text{for} \qquad k - n < -1.
\end{equation}
In light of the aforementioned considerations, it is possible to rewrite eq. \eqref{eq:168} in the following form:
\begin{equation}
\label{eq:170}
\Lambda^{-1}\left( \Lambda \frac{\partial }{\partial \Lambda}  \right)^l \left( \Lambda^{-1}\left( \Lambda \frac{\partial }{\partial \Lambda}  \right)^m  \right)^{n - 1} f(\mathbf{p})= \sum_{k \ge - 1}^{}\Lambda^k\widetilde{W}_{k + n}^{(-m, -n | -l)} f(\mathbf{p}). 
\end{equation}
After moving \(\Lambda^{-1}\) to the right-hand side
\begin{dmath}
\label{eq:172}
\left( \Lambda \frac{\partial }{\partial \Lambda}  \right)^l \left( \Lambda^{-1}\left( \Lambda \frac{\partial }{\partial \Lambda}  \right)^m  \right)^{n - 1} f(\mathbf{p})= \sum_{k \ge - 1}^{}\Lambda^{k + 1}\widetilde{W}_{k + n}^{(-m, -n | -l)} f(\mathbf{p}) =  \sum_{k \ge 0}^{}\Lambda^{k}\widetilde{W}_{k + n - 1}^{(-m, -n | -l)} f(\mathbf{p}), 
\end{dmath}
In other words
\begin{equation}
\label{eq:194}
\left( \Lambda \frac{\partial }{\partial \Lambda}  \right)^l \left( \Lambda^{-1}\left( \Lambda \frac{\partial }{\partial \Lambda}  \right)^m  \right)^{n} f(\mathbf{p}) =  \sum_{k \ge 0}^{}\Lambda^{k}\widetilde{W}_{k + n}^{(-m, -n -1 | -l)} f(\mathbf{p}).
\end{equation}
In this form, it is evident that the desired \(\widetilde{W}_k^{(-m, -n)}\)-s are merely a special case of the auxiliary operators
\begin{equation}
\label{eq:200}
\widetilde{W}_k^{(-m, -n)} = \widetilde{W}_k^{(-m, -n - 1|0)}.
\end{equation}
The base of the recursion can be found by considering the \(l = 0\) \& \(n = 1\) case
\begin{equation}
\label{eq:174}
\widetilde{W}_k^{(-m, -1| 0)} = \delta_{k,0}.
\end{equation}
In the case of \(l = m\) we have
\begin{equation}
\label{eq:175}
\left( \Lambda^{-1} \left( \Lambda \frac{\partial }{\partial \Lambda}  \right)^m \right)^n f(\mathbf{p})= \sum_{k \ge -1}^{}\Lambda^k \widetilde{W}_{k + n}^{(-m, -n | -m)} f(\mathbf{p}),
\end{equation}
while for \(l = 0\)
\begin{equation}
\label{eq:176}
\left( \Lambda^{-1}\left( \Lambda \frac{\partial }{\partial \Lambda}  \right)^m  \right)^{n} f(\mathbf{p})= \sum_{k \ge 0}^{}\Lambda^{k}\widetilde{W}_{k + n}^{(-m, -n - 1| 0)} f(\mathbf{p}).
\end{equation}
Therefore,
\begin{equation}
\label{eq:177}
\widetilde{W}_k^{(-m, -n - 1| 0)} = \widetilde{W}_k^{(-m, -n, | -m)} \qquad \text{for} \qquad k - n \ge 0,
\end{equation}
\begin{equation}
\label{eq:193}
\widetilde{W}_k^{(-m, -n - 1| 0)} = 0 \qquad \text{otherwise}.
\end{equation}
Next, the step of the recursion in \(l\) direction should be evaluated. The chain of equalities for that problem is as follows
\begin{equation}
\label{eq:197}
\begin{aligned}
\sum_{k \ge 0}^{} \Lambda^k \widetilde{W}_{k + n - 1}^{(-m, -n| -l - 1)} f(\mathbf{p}) &=\left( \Lambda \frac{\partial }{\partial \Lambda}  \right)^{l + 1} \left( \Lambda^{-1} \left( \Lambda \frac{\partial }{\partial \Lambda}  \right)^m \right)^{n - 1} f(\mathbf{p}) \\ & = \left( \Lambda \frac{\partial }{\partial \Lambda}  \right) \left( \Lambda \frac{\partial }{\partial \Lambda}  \right)^{l} \left( \Lambda^{-1} \left( \Lambda \frac{\partial }{\partial \Lambda} \right)^m \right)^{n - 1} f(\mathbf{p}) \\ & = \Lambda \frac{\partial }{\partial \Lambda} \sum_{k \ge 0}^{}\Lambda^k \widetilde{W}_{k + n - 1}^{(-m, -n| - l)} f(\mathbf{p}) \\ & =
\sum_{k \ge 0}^{} \left( \sum_{r = 0}^{k - 1}\Lambda^{k - r}p_r  + \sum_{r \ge 1}^{}r \Lambda^{k + r} \frac{\partial }{\partial p_r}  \right) \widetilde{W}_{k + n -1}^{(-m, -n | -l)} f(\mathbf{p}) \\ & =
\sum_{k \ge 1}^{} \Lambda^k \left( \sum_{r \ge 0}^{} p_r \widetilde{W}_{k + n + r - 1}^{(-m, -n| -l)} +
\sum_{r = 1}^{k} r \frac{\partial }{\partial p_r} \widetilde{W}_{k + n - r - 1}^{(-m, -n | -l)}  \right) f(\mathbf{p}).
\end{aligned}
\end{equation}
A comparison of the leftmost side with the rightmost shows that
\begin{equation}
\label{eq:198}
\widetilde{W}_k^{(-m, -n | - l - 1)} = \sum_{r \ge 0}^{} \widetilde{W}_{k + r}^{(-m, -n| -l)} +
\sum_{r = 1}^{k - n + 1} r \frac{\partial }{\partial p_r} \widetilde{W}_{k - r}^{(-m, -n | -l)} \qquad \text{for} \qquad k - n \ge 0,
\end{equation}
\begin{equation}
\label{eq:199}
\widetilde{W}_k^{(-m, -n | -l - 1)} = 0 \qquad \text{otherwise}.
\end{equation}
The collective set of all the ingredients that comprise the obtained recursion is as follows:
\begin{subequations}
\label{eq:200}
\begin{empheq}[box=\fbox]{gather}
\label{eq:201}
\widetilde{W}_k^{(-m, -n)} = \widetilde{W}_k^{(-m, -n - 1 | 0)}, \\
\label{eq:202}
\begin{aligned}
\widetilde{W}_{k}^{(-m, -n | -l - 1)}  = \sum_{r \ge 0}^{}p_r \widetilde{W}_{k + r}^{(-m, -n |-l)}
+ \sum_{r = 1}^{k - n + 1} r \frac{\partial }{\partial p_r}  \widetilde{W}_{k - r}^{(-m, -n |-l)} & \\
\qquad \text{for}\qquad k - n \ge 0,&
\end{aligned} \\
\label{eq:203}
  \widetilde{W}_{k}^{(-m, -n - 1| 0)} = \widetilde{W}_{k}^{(-m, -n | -m)} \qquad \text{for} \qquad n > 0,\\
\label{eq:204}
\widetilde{W}_k^{(-m, -1 | 0)} = \delta_{k, 0}, \\
\widetilde{W}_k^{(-m, -n | -l)} = 0 \qquad \text{otherwise.} 
\end{empheq}
\end{subequations}
It is necessary to perform a sanity check. Let us begin by demonstrating the equivalence of the \(\widetilde{W}_k^{(m, -)}\) and the \(\widetilde{W}_k^{(-m, -1)}\) operators. First step of the evaluation of \(\widetilde{W}_k^{(-m, -1)}\)
\begin{equation}
\label{eq:201}
\widetilde{W}_k^{(-m, -1)} = \widetilde{W}_k^{(-m, -2|0)} = \widetilde{W}_k^{(-m, -1| -m)}
\end{equation}
is based on the application of eqs. \eqref{eq:201} and \eqref{eq:203}. Subsequently, the recursion of \eqref{eq:202}
\begin{equation}
\label{eq:165}
\widetilde{W}_k^{(-m, -1| -l -1)} =  \sum_{r \ge 0}^{}p_r \widetilde{W}_{k + r}^{(-m, -1 |-l)}
+ \sum_{r = 1}^{k} r \frac{\partial }{\partial p_r}  \widetilde{W}_{k - r}^{(-m, -1 |-l)}
\end{equation}
follows for \(k \ge 1\) with the base given by \eqref{eq:204}. This is precisely the recursion described in eqs. \eqref{eq:144}, \eqref{eq:153}, \eqref{eq:154}.

The case of the \(\widetilde{W}^{(-, n)}\) algebra is also worthy of consideration here. In this case, we have for \(k - n \ge 1\)
\begin{dmath}
\label{eq:196}
\widetilde{W}_k^{(-1, -n - 1)} = \widetilde{W}_k^{(-1, -n - 2|  0)}= \widetilde{W}_k^{(-1, -n -1 | -1)} = \sum_{r \ge 0}^{} p_r \widetilde{W}_{k + r}^{(-1, -n - 1| 0)} + \sum_{r = 1}^{k - n} r \frac{\partial }{\partial p_r} \widetilde{W}_{k- r}^{(-1, -n - 1 |0)} = \sum_{r \ge 0}^{} p_r \widetilde{W}_{k + r}^{(-1, -n)} + \sum_{r \ge 0}^{k - n} r \frac{\partial }{\partial p_r} \widetilde{W}_{k - r}^{(-1, -n)}
\end{dmath}
with the base given by \eqref{eq:204}
\begin{equation}
\label{eq:204}
\widetilde{W}_k^{(-1,0)} = \widetilde{W}^{(-1,1|0)}_k= \delta_{k, 0}.
\end{equation}
It is the anticipated recursion procedure of the \(\widetilde{W}^{(-, n)}\) algebra. 
\subsection{Ward identities}
\label{sec:org8308eb9}
As previously stated in \eqref{eq:7}, the partition functions \(Z_n\) satisfy the following set of \(\widetilde{W}^{(1,n)}\) constraints \cite{mironov-2023-spect-curves,alexandrov-2014-kp-integ}
\begin{equation}
\widetilde{W}_k^{(n)} Z_{n} = (k + n)
\frac{\partial Z_n}{\partial p_{k + n}}, \qquad k + n \ge 1. 
\end{equation}
All of them can be encapsulated within a single equation \cite{mironov-2021-viras-versus-super,mironov-2021-matrix-model,mironov-2021-non-abelian}
\begin{equation}
\label{eq:247}
\Bigg( \sum_{k \ge 1}^{} k p_k \frac{\partial }{\partial p_k}  - \underbrace{\sum_{k \ge 0}^{} p_k \widetilde{W}_{k - n}^{(1, n)}}_{H_n^{(1)}} \Bigg) Z_n = 0
\end{equation}
with the solution known to be
\begin{equation}
\label{eq:248}
Z_{n} = e^{ \frac{1}{n} H_n^{(1)}}\cdot 1.
\end{equation} 
This type of reasoning can be applied to the generalized case of
\begin{equation}
\label{eq:249}
Z_n^{(m)} = e^{\frac{1 }{n} H_n^{(m)}} \cdot 1.
\end{equation}
In particular, \(Z_n^{(m)}\) satisfies a single equation
\begin{equation}
\label{eq:250}
\Bigg( \sum_{k \ge 1}^{} k p_k \frac{\partial }{\partial p_k}  - \underbrace{\sum_{k \ge 0}^{} p_k \widetilde{W}_{k - n}^{(m, n)}}_{H_n^{(m)}} \Bigg) Z_n^{(m)} = 0.
\end{equation}
Not only this fact, but also rich computational evidence allows us to make a conjecture that Ward identities for \(Z_n^{(m)}\)are as follows:
\begin{equation}
\label{eq:64}
\boxed{\widetilde{W}_n^{(m, n)} Z_{n}^{(m)} = (n + k) \frac{\partial Z_{n}^{(m)}}{\partial p_{n + k}}, \qquad k + n \ge 1. }    
\end{equation}
\section{Vertical ray Hamiltonians and related}
\label{sec:org6fde181}
The distinguished Hamiltonians on the vertical ray necessitate a specific approach. Subsequently, one can proceed to the investigation of the \(\widetilde{W}\) algebras, situated on this ray. In the first part of this section, the general framework necessary for the subsequent computations is introduced. Next, the sought-after expression of Hamiltonians on the vertical ray is derived and, as a demonstration of the tool's utility, some matrix \(W_{1+\infty}\) identities are demonstrated to be true using this approach.
\subsection{Matrix Heisenberg algebra and general framework}
\label{sec:org3ea4a67}
The simplest objects made of matrix elements that have a non-trivial commutator are
\begin{dmath}
\label{eq:mc1}
\left( \frac{\partial }{\partial \Lambda}  \right)_{ij} \Lambda_{kl} =  \delta_{il}\delta_{jk} + \Lambda_{kl}\left( \frac{\partial }{\partial \Lambda}  \right)_{ij}, 
\end{dmath}
so
\begin{dmath}
\label{eq:mc2}
\left[ \left( \frac{\partial }{\partial \Lambda}  \right)_{ij}, \Lambda_{kl} \right] = \delta_{il} \delta_{jk}. 
\end{dmath}
Of course, \(\Lambda_{ij}\)-s commute between themselves, as well as derivatives by them
\begin{equation}
\label{eq:mc6}
\left[ \Lambda_{ij}, \Lambda_{kl} \right] = 0, \qquad \left[ \frac{\partial }{\partial \Lambda_{ij}}, \frac{\partial }{\partial \Lambda_{kl}}   \right] = 0. 
\end{equation}
On the next level of complexity, we have commutators
\begin{dmath}
\label{eq:mc3}
\left[ \left( \frac{\partial }{\partial \Lambda} \Lambda \right)_{ij}, \Lambda_{kl} \right] = \left[ \left( \frac{\partial }{\partial \Lambda}  \right)_{in}\Lambda_{nj}, \Lambda_{kl} \right] = \left[ \left( \frac{\partial }{\partial \Lambda}  \right)_{in}, \Lambda_{kl} \right] \Lambda_{nj} = \delta_{il}\delta_{nk}\Lambda_{nj}= \delta_{il} \Lambda_{kj} 
\end{dmath}
and
\begin{dmath}
\label{eq:mc4}
\left[ \left( \Lambda\frac{\partial }{\partial \Lambda}  \right)_{ij}, \Lambda_{kl} \right] = \Lambda_{il} \delta_{kj},  
\end{dmath}
as well as
\begin{dmath}
\label{eq:mc7}
\left[ \left( \frac{\partial }{\partial \Lambda}  \right)_{ij}, \left( \frac{\partial }{\partial \Lambda} \Lambda \right)_{kl}  \right] = \left( \frac{\partial }{\partial \Lambda}  \right)_{kn} \left[ \left( \frac{\partial }{\partial \Lambda}  \right)_{ij} , \Lambda_{nl}\right] = \left( \frac{\partial }{\partial \Lambda}  \right)_{kn}\delta_{il}\delta_{jn} = \delta_{il} \left( \frac{\partial }{\partial \Lambda}  \right)_{kj} 
\end{dmath}
and
\begin{dmath}
\label{eq:mc15}
\left[ \left( \frac{\partial }{\partial \Lambda}  \right)_{ij}, \left( \Lambda \frac{\partial }{\partial \Lambda}  \right)_{kl} \right] = \delta_{jk} \left( \frac{\partial }{\partial \Lambda}  \right)_{il}.  
\end{dmath}
It turns out that the following zero-grading elements commute between themselves
\begin{dmath}
\label{eq:mc5}
\left[ \left( \frac{\partial }{\partial \Lambda} \Lambda \right)_{ij}, \left( \Lambda \frac{\partial }{\partial \Lambda}  \right)_{kl} \right] = 0. 
\end{dmath}
It can be straightforwardly proven using aforementioned commutators
\begin{dmath}
\left( \frac{\partial }{\partial \Lambda} \Lambda \right)_{ij} \left( \Lambda \frac{\partial }{\partial \Lambda}  \right)_{kl} =  \left( \frac{\partial }{\partial \Lambda}  \right)_{in} \Lambda_{nj}  \left( \Lambda \frac{\partial }{\partial \Lambda}  \right)_{kl} =  \left( \frac{\partial }{\partial \Lambda}  \right)_{in} \left(   \left( \Lambda \frac{\partial }{\partial \Lambda}  \right)_{kl} \Lambda_{nj} - \delta_{nl}\Lambda_{kj}\right) = \left( \left( \Lambda \frac{\partial }{\partial \Lambda}  \right)_{kl} \left( \frac{\partial }{\partial \Lambda}  \right)_{in} \Lambda_{nj} + \delta_{kn}\left( \frac{\partial }{\partial \Lambda}  \right)_{il} \Lambda_{nj}\right) - \left( \frac{\partial }{\partial \Lambda}  \right)_{il} \Lambda_{kj} =\left( \Lambda \frac{\partial }{\partial \Lambda}  \right)_{kl} \left( \frac{\partial }{\partial \Lambda} \Lambda \right)_{ij} .  
\end{dmath}
This is a particularly important tool for the calculations below. Let us also introduce some natural definitions
\begin{equation}
\label{eq:mc12}
\mathcal{E}_m = \left( \Lambda \frac{\partial }{\partial \Lambda}  \right)^m\Lambda, \qquad \mathcal{F}_{m} = \Lambda^{-1} \left( \frac{\partial }{\partial \Lambda} \Lambda \right)^m.
\end{equation}
The names for these operators have been chosen to correspond to the traced operators
\begin{equation}
\label{eq:211}
E_m = \tr \mathcal{E}_m, \qquad F_m = \tr \mathcal{F}_m.
\end{equation}
It is also worth mentioning that even WLZZ Hamiltonians can be easily rewritten through them
\begin{equation}
\label{eq:212}
H_n^{(m)} = \tr \mathcal{E}_m^n, \qquad H_{-n}^{(-m)} = \tr \mathcal{F}_m^n.
\end{equation}
The defining characteristics of such operators are as follows:
\begin{equation}
\mathcal{F}_{n + 1} = \mathcal{F}_n \left( \Lambda \frac{\partial }{\partial \Lambda}  \right) = \left( \frac{\partial }{\partial \Lambda} \Lambda \right) \mathcal{F}_n, \qquad \mathcal{E}_{m + 1} = \mathcal{E}_m \left( \frac{\partial }{\partial \Lambda}  \Lambda \right) = \left( \Lambda \frac{\partial }{\partial \Lambda}  \right)\mathcal{E}_{m}
\end{equation}
It is important to consider the commutators of such operators before moving on to more complex examples
\begin{dmath}
\label{eq:mc10}
\left[ \left( \frac{\partial }{\partial \Lambda} \Lambda \right)_{ij}, \left( \mathcal{E}_m \right)_{kl}  \right] =  \left[ \left( \frac{\partial }{\partial \Lambda} \Lambda \right)_{ij}, \left( \Lambda \frac{\partial }{\partial \Lambda}  \right)_{kn}^{m}  \Lambda_{nl}  \right] = \left( \Lambda \frac{\partial }{\partial \Lambda}  \right)_{kn}^m \left[ \left( \frac{\partial }{\partial \Lambda} \Lambda \right)_{ij},  \Lambda_{nl}  \right]= \left( \Lambda \frac{\partial }{\partial \Lambda}  \right)_{kn}^{m} \delta_{il} \Lambda_{nj} = \delta_{il} \left( \mathcal{E}_m \right)_{kj} ,
\end{dmath}
\begin{dmath}
\left[ \left( \mathcal{F}_m \right)_{kl},\left(  \frac{\partial }{\partial \Lambda} \Lambda \right)_{ij} \right] = \left[  \left( \frac{\partial }{\partial \Lambda} \right)_{kn} \left( \Lambda \frac{\partial }{\partial \Lambda}  \right)_{nl}^{m - 1} , \left(  \frac{\partial }{\partial \Lambda} \Lambda \right)_{ij}\right] = \left[ \left( \frac{\partial }{\partial \Lambda}  \right)_{kn},\left(  \frac{\partial }{\partial \Lambda} \Lambda \right)_{ij}  \right] \left( \Lambda \frac{\partial }{\partial \Lambda}  \right)_{nl}^{m - 1} =   \delta_{kj}\left( \frac{\partial }{\partial \Lambda}  \right)_{in} \left( \Lambda \frac{\partial }{\partial \Lambda}  \right)_{nl}^{m - 1} = \delta_{kj} \left( \mathcal{F}_m \right)_{il}.
\end{dmath}
Analogously
\begin{dmath}
\label{eq:mc11}
\left[ \left( \Lambda \frac{\partial }{\partial \Lambda}  \right)_{ij}, \left( \mathcal{E}_m\right)_{kl} \right] = \delta_{kj} \left(\mathcal{E}_m\right)_{il}
\end{dmath}
and
\begin{dmath}
\label{eq:mc11}
\left[ \left( \mathcal{F}_m\right)_{kl},\left( \Lambda \frac{\partial }{\partial \Lambda}  \right)_{ij} \right] = \delta_{il} \left(\mathcal{F}_m\right)_{kj}.
\end{dmath}
\subsection{Finding the expression for \(\left[ F_n, E_m \right]\)}
\label{sec:orgaaface4}
To gain insight into the elements of zero grading, one can consider the commutator of the elements of grading 1 and \(-1\)
\begin{dmath}
\label{eq:mc14}
\left[ F_n, E_m \right] = F_n E_m -E_m F_n.
\end{dmath}
The first term of the commutator in \eqref{eq:mc14} is, in fact,
\begin{dmath}
F_n E_m = \tr \mathcal{F}_n \tr \mathcal{E}_{m},
\end{dmath}
which can be rewritten in index notation as
\begin{dmath}
F_n E_m = \left( \mathcal{F}_n \right)_{ii} \left( \mathcal{E}_m \right)_{jj}. 
\end{dmath}
To make use of commutator \eqref{eq:mc11} one should separate the last \(\Lambda  \frac{\partial }{\partial \Lambda}\) from the \(\mathcal{F}_n\) (if \(n\) is greater than zero, of course)
\begin{dmath}
F_n E_m = \left( \mathcal{F}_{n - 1} \right)_{ik} \left( \Lambda \frac{\partial }{\partial \Lambda}  \right)_{ki} \left( \mathcal{E}_m\right)_{jj}.
\end{dmath}
We can now move \(\Lambda \frac{\partial }{\partial \Lambda}\) to the right
\begin{dmath}
F_n E_m = \left( \mathcal{F}_{n - 1} \right)_{ik} \left(  \left( \mathcal{E}_m\right)_{jj} \left( \Lambda \frac{\partial }{\partial \Lambda}  \right)_{ki} + \delta_{ij} \left( \mathcal{E}_m \right)_{kj}\right).
\end{dmath}
The expansion of this expression reads as follows
\begin{dmath}
F_n E_m = \left( \mathcal{F}_{n - 1} \right)_{ik}  \left( \mathcal{E}_m\right)_{jj} \left( \Lambda \frac{\partial }{\partial \Lambda}  \right)_{ki} + \left( \mathcal{F}_{n - 1} \mathcal{E}_m \right)_{jj}.
\end{dmath}
The second term is
\begin{dmath}
\mathcal{F}_{n - 1} \mathcal{E}_m = \Lambda^{-1} \left( \Lambda \frac{\partial }{\partial \Lambda}  \right)^{n - 1} \Lambda \left( \frac{\partial }{\partial \Lambda}  \Lambda\right)^m = \left( \frac{\partial }{\partial \Lambda} \Lambda \right)^{m + n - 1} .
\end{dmath}
Let us repeat this cycle once again
\begin{dmath}
F_n E_m = \left( \mathcal{F}_{n - 2} \right)_{il}  \left( \Lambda \frac{\partial }{\partial \Lambda}  \right)_{lk}\left( \mathcal{E}_m\right)_{jj} \left( \Lambda \frac{\partial }{\partial \Lambda}  \right)_{ki} + \tr \left( \frac{\partial }{\partial \Lambda} \Lambda \right)^{m + n - 1} = \left( \mathcal{F}_{n - 2} \right)_{il}  \left( \left( \mathcal{E}_m\right)_{jj} \left( \Lambda \frac{\partial }{\partial \Lambda}  \right)_{lk} + \delta_{kj} \left( \mathcal{E}_{m} \right)_{lj}\right) \left( \Lambda \frac{\partial }{\partial \Lambda}  \right)_{ki} + \tr \left( \frac{\partial }{\partial \Lambda} \Lambda \right)^{m + n - 1} = \left( \mathcal{F}_{n - 2} \right)_{il}   \left( \mathcal{E}_m\right)_{jj} \left( \Lambda \frac{\partial }{\partial \Lambda}  \right)_{lk}  \left( \Lambda \frac{\partial }{\partial \Lambda}  \right)_{ki} + \left( \mathcal{F}_{n - 2} \right)_{il}    \left( \mathcal{E}_m \right)_{lk} \left( \Lambda \frac{\partial }{\partial \Lambda}  \right)_{ki} + \tr \left( \frac{\partial }{\partial \Lambda} \Lambda \right)^{m + n - 1} = \left( \mathcal{F}_{n - 2} \right)_{ik}   \left( \mathcal{E}_m\right)_{jj} \left( \Lambda \frac{\partial }{\partial \Lambda}  \right)_{ki}^{2}   + \left( \mathcal{F}_{n - 2} \mathcal{E}_m  \left( \Lambda \frac{\partial }{\partial \Lambda} \right)  \right)_{ii} + \tr \left( \frac{\partial }{\partial \Lambda} \Lambda \right)^{m + n - 1} = \left( \mathcal{F}_{n - 2} \right)_{ik}   \left( \mathcal{E}_m\right)_{jj} \left( \Lambda \frac{\partial }{\partial \Lambda}  \right)_{ki}^{2}   + \tr\left( \left( \frac{\partial }{\partial \Lambda} \Lambda \right)^{m + n - 2}  \left( \Lambda \frac{\partial }{\partial \Lambda} \right)  \right) + \tr \left( \frac{\partial }{\partial \Lambda} \Lambda \right)^{m + n - 1}
\end{dmath}
Upon reaching the \(l\)-th step of the aforementioned procedure, the following expression will be obtained:
\begin{dmath}
F_n E_m =  \left( \mathcal{F}_{n - l} \right)_{ik}   \left( \mathcal{E}_m\right)_{jj} \left( \Lambda \frac{\partial }{\partial \Lambda}  \right)_{ki}^{l }   + \sum_{r = 1}^{l}\tr\left( \left( \frac{\partial }{\partial \Lambda} \Lambda \right)^{m + n - r}  \left( \Lambda \frac{\partial }{\partial \Lambda} \right)^{r - 1}  \right).
\end{dmath}
We should stop at the \(n - 1\)-th step, for which we have
\begin{dmath}
\label{eq:mc16}
F_n E_m =  \left( \frac{\partial }{\partial \Lambda}  \right)_{ik}   E_m \left( \Lambda \frac{\partial }{\partial \Lambda}  \right)_{ki}^{n - 1}   + \sum_{r = 1}^{n - 1 }\tr\left( \left( \frac{\partial }{\partial \Lambda} \Lambda \right)^{m + n - r}  \left( \Lambda \frac{\partial }{\partial \Lambda} \right)^{r - 1}  \right) .\end{dmath}
The next step is to utilize the commutator \eqref{eq:mc15} to derive the expression for
\begin{dmath}
\left( \frac{\partial }{\partial \Lambda}  \right)_{ij} E_m  =  \left( \frac{\partial }{\partial \Lambda}  \right)_{ij} \left( \Lambda \frac{\partial }{\partial \Lambda}  \right)_{kl} \left( \mathcal{E}_{m - 1} \right)_{lk} =  \left( \left( \Lambda \frac{\partial }{\partial \Lambda}  \right)_{kl}\left( \frac{\partial }{\partial \Lambda}  \right)_{ij} + \delta_{jk} \left( \frac{\partial }{\partial \Lambda}  \right)_{il} \right)\left( \mathcal{E}_{m - 1} \right)_{lk} = \left( \Lambda \frac{\partial }{\partial \Lambda}  \right)_{kl}\left( \frac{\partial }{\partial \Lambda}  \right)_{ij} \left( \mathcal{E}_{m - 1} \right)_{lk} +   \left( \frac{\partial }{\partial \Lambda}  \right)_{il} \left( \mathcal{E}_{m - 1} \right)_{lj} = \left( \Lambda \frac{\partial }{\partial \Lambda}  \right)_{kl}\left( \frac{\partial }{\partial \Lambda}  \right)_{ij} \left( \mathcal{E}_{m - 1} \right)_{lk} +   \left( \frac{\partial }{\partial \Lambda} \Lambda \right)_{ij}^m.
\end{dmath}
Once more, a single step was insufficient; it would be beneficial to observe the situation at the second step
\begin{dmath}
\left( \frac{\partial }{\partial \Lambda}  \right)_{ij} E_m  = \left( \Lambda \frac{\partial }{\partial \Lambda}  \right)_{kl}\left( \frac{\partial }{\partial \Lambda}  \right)_{ij} \left( \Lambda \frac{\partial }{\partial \Lambda}  \right)_{lp}\left( \mathcal{E}_{m - 2} \right)_{pk} +   \left( \frac{\partial }{\partial \Lambda} \Lambda \right)_{ij}^m = \left( \Lambda \frac{\partial }{\partial \Lambda}  \right)_{kl}\left(  \left( \Lambda \frac{\partial }{\partial \Lambda}  \right)_{lp} \left( \frac{\partial }{\partial \Lambda}  \right)_{ij} + \delta_{jl}\left( \frac{\partial }{\partial \Lambda}  \right)_{ip}\right)\left( \mathcal{E}_{m - 2} \right)_{pk} +   \left( \frac{\partial }{\partial \Lambda} \Lambda \right)_{ij}^m = \left( \Lambda \frac{\partial }{\partial \Lambda}  \right)_{kp}^2 \left( \frac{\partial }{\partial \Lambda}  \right)_{ij} \left( \mathcal{E}_{m - 2} \right)_{pk} +\left( \Lambda \frac{\partial }{\partial \Lambda}  \right)_{kj}    \left( \frac{\partial }{\partial \Lambda}  \right)_{ip}\left( \mathcal{E}_{m - 2} \right)_{pk} +   \left( \frac{\partial }{\partial \Lambda} \Lambda \right)_{ij}^m = \left( \Lambda \frac{\partial }{\partial \Lambda}  \right)_{kp}^2 \left( \frac{\partial }{\partial \Lambda}  \right)_{ij} \left( \mathcal{E}_{m - 2} \right)_{pk} +\left( \Lambda \frac{\partial }{\partial \Lambda}  \right)_{kj}    \left( \frac{\partial }{\partial \Lambda}\Lambda  \right)^{m - 1}_{ik} +   \left( \frac{\partial }{\partial \Lambda} \Lambda \right)_{ij}^m.
\end{dmath}
The general structure can now be understood (and, of course, proven by induction). On the \(l\)-s step, one can get
\begin{dmath}
\left(  \frac{\partial }{\partial \Lambda} \right)_{ij} E_m = \left( \Lambda \frac{\partial }{\partial \Lambda}  \right)_{kp}^l \left( \frac{\partial }{\partial \Lambda}  \right)_{ij} \left( \mathcal{E}_{m - l} \right)_{pk} + \sum_{r = 0}^{l - 1}\left( \Lambda \frac{\partial }{\partial \Lambda}  \right)_{kj}^{r} \left( \frac{\partial }{\partial \Lambda} \Lambda \right)_{ik}^{m - r}.
\end{dmath}
The entire process will end at the \(m\)-s step, resulting in
\begin{dmath}
\left(  \frac{\partial }{\partial \Lambda} \right)_{ij} E_m = \left( \Lambda \frac{\partial }{\partial \Lambda}  \right)_{kp}^m   \left( \frac{\partial }{\partial \Lambda}  \right)_{ij} \Lambda_{pk} + \sum_{r = 0}^{m - 1}\left( \Lambda \frac{\partial }{\partial \Lambda}  \right)_{kj}^{r} \left( \frac{\partial }{\partial \Lambda} \Lambda \right)_{ik}^{m - r}. 
\end{dmath}
Finally, the \(\frac{\partial }{\partial \Lambda}\) and \(\Lambda\) should be commuted
\begin{dmath}
\left(  \frac{\partial }{\partial \Lambda} \right)_{ij} E_m = \left( \Lambda \frac{\partial }{\partial \Lambda}  \right)_{kp}^m   \left( \Lambda_{pk} \left( \frac{\partial }{\partial \Lambda}  \right)_{ij} + \delta_{ik}\delta_{pj} \right) + \sum_{r = 0}^{m - 1}\left( \Lambda \frac{\partial }{\partial \Lambda}  \right)_{kj}^{r} \left( \frac{\partial }{\partial \Lambda} \Lambda \right)_{ik}^{m - r} =E_m \left( \frac{\partial }{\partial \Lambda}  \right)_{ij}  + \sum_{r = 0}^{m}\left( \Lambda \frac{\partial }{\partial \Lambda}  \right)_{kj}^{r} \left( \frac{\partial }{\partial \Lambda} \Lambda \right)_{ik}^{m - r}. 
\end{dmath}
This can be pasted in \eqref{eq:mc16}
\begin{dmath}
F_n E_m =  \left( \frac{\partial }{\partial \Lambda}  \right)_{ik}   E_m \left( \Lambda \frac{\partial }{\partial \Lambda}  \right)_{ki}^{n - 1}   + \sum_{r = 1}^{n - 1 }\tr\left( \left( \frac{\partial }{\partial \Lambda} \Lambda \right)^{m + n - r}  \left( \Lambda \frac{\partial }{\partial \Lambda} \right)^{r - 1}  \right)   = \left( E_m \left( \frac{\partial }{\partial \Lambda}  \right)_{ik}  + \sum_{r = 0}^{m}\left( \Lambda \frac{\partial }{\partial \Lambda}  \right)_{jk}^{r} \left( \frac{\partial }{\partial \Lambda} \Lambda \right)_{ij  }^{m - r}  \right) \left( \Lambda \frac{\partial }{\partial \Lambda}  \right)_{ki}^{n - 1}   + \sum_{r = 1}^{n - 1 }\tr\left( \left( \frac{\partial }{\partial \Lambda} \Lambda \right)^{m + n - r}  \left( \Lambda \frac{\partial }{\partial \Lambda} \right)^{r - 1}  \right) = E_m F_n +  \sum_{r = 0}^{m}\tr \left( \left( \frac{\partial }{\partial \Lambda} \Lambda \right)^{m - r } \left( \Lambda \frac{\partial }{\partial \Lambda}  \right)^{r + n - 1}  \right) + \sum_{r = 1}^{n - 1 }\tr\left( \left( \frac{\partial }{\partial \Lambda} \Lambda \right)^{m + n - r}  \left( \Lambda \frac{\partial }{\partial \Lambda} \right)^{r - 1}  \right) =  E_m F_n + \sum_{r = 1}^{ m + n} \tr\left( \left( \frac{\partial }{\partial \Lambda} \Lambda \right)^{m + n - r}  \left( \Lambda \frac{\partial }{\partial \Lambda} \right)^{r - 1}  \right).
\end{dmath}
The conclusion is that
\begin{equation}
\label{eq:213}
\left[ F_n, E_m \right] = H_{0}^{(m + n - 1)} = \sum_{r = 1}^{ m + n} \tr\left( \left( \frac{\partial }{\partial \Lambda} \Lambda \right)^{m + n - r}  \left( \Lambda \frac{\partial }{\partial \Lambda} \right)^{r - 1}  \right).
\end{equation}
\subsection{Proof of the recursion relation for \(E_m\)}
\label{sec:org04de9ad}
The goal of this section is to prove the relation
\begin{dmath}
\label{eq:em1}
E_{m + 1} = \left[ W_0, E_m \right],  
\end{dmath}
where
\begin{dmath}
\label{eq:206}
W_0= \frac{1}{2} \tr \left( \frac{\partial }{\partial \Lambda} \Lambda^2 \frac{\partial }{\partial \Lambda}  \right) = \frac{1}{2} \left( \frac{\partial }{\partial \Lambda} \Lambda \right)_{ij} \left( \Lambda \frac{\partial }{\partial \Lambda}  \right)_{ji}. 
\end{dmath}
The first term on the right-hand side of \eqref{eq:em1} is expressed as
\begin{dmath}
W_0 E_m = \frac{1}{2} \left( \frac{\partial }{\partial \Lambda} \Lambda \right)_{ij} \left( \Lambda \frac{\partial }{\partial \Lambda}  \right)_{ji}  \left( \mathcal{E}_m \right)_{kk} = \frac{1}{2}  \left( \Lambda \frac{\partial }{\partial \Lambda}  \right)_{ji} \left( \frac{\partial }{\partial \Lambda} \Lambda \right)_{ij} \left( \mathcal{E}_m \right)_{kk}.  
\end{dmath}
The commutation relation \eqref{eq:mc11} yields
\begin{dmath}
W_0 E_m = \frac{1}{2} \left( \Lambda \frac{\partial }{\partial \Lambda}  \right)_{ji} \left( \left( \mathcal{E}_m \right)_{kk} \left(  \frac{\partial }{\partial \Lambda}\Lambda  \right)_{ij} + \delta_{ik} \left( \mathcal{E}_m \right)_{kj}\right) =  \frac{1}{2} \left( \tr \left( \Lambda \frac{\partial }{\partial \Lambda} E_m \frac{\partial }{\partial \Lambda} \Lambda \right) +  E_{m + 1} \right).  
\end{dmath}
Similarly,
\begin{dmath}
E_m W_0 = \frac{1}{2} \left( \tr \left( \Lambda \frac{\partial }{\partial \Lambda} E_m \frac{\partial }{\partial \Lambda} \Lambda \right) -  E_{m + 1} \right).  
\end{dmath}
Thus,
\begin{dmath}
\left[ W_0, E_m \right] =W_0 E_m - E_{m }W_0 =E_{m + 1}.   
\end{dmath}
\subsection{Proof of the recursion relation for \(H_n^{(m)}\)}
\label{sec:org322b365}
This section is devoted to proving the fact that
\begin{dmath}
n H_{n + 1}^{(m)} = \left[ E_{m + 1}, H_n^{(m)} \right]. 
\end{dmath}
As a starting point, we have
\begin{dmath}
\left[ E_{m + 1}, H_n^{(m)} \right] = E_{m + 1} H_n^{(m)} - H_n^{(m)} E_{m + 1} .  
\end{dmath}
One final auxiliary step is necessary to advance. It is necessary to demonstrate that
\begin{dmath}
\left[\left( \mathcal{E}_{m + 1}\right)
_{ij}, \left( \mathcal{E}_m \right)_{kl} \right] = \left( \mathcal{E}_m\right)_{kj} \left( \mathcal{E}_m\right)_{il}  
\end{dmath}
by induction. The base of the induction is
\begin{dmath}
\left[ \left( \Lambda \frac{\partial }{\partial \Lambda} \Lambda \right)_{ij} , \Lambda_{kl} \right] = \left[ \Lambda_{in}\left(  \frac{\partial }{\partial \Lambda} \Lambda \right)_{nj} , \Lambda_{kl} \right] = \Lambda_{in}\left[ \left(  \frac{\partial }{\partial \Lambda} \Lambda \right)_{nj} , \Lambda_{kl} \right] = \Lambda_{in} \Lambda_{kj} \delta_{ln}= \Lambda_{il} \Lambda_{kj}, 
\end{dmath}
Next, the step of the recursion follows. The objective is to reduce this expression to a form consisting of the commutator \(\left[ \left( \mathcal{E}_{m} \right)_{ij}, \left( \mathcal{E}_{m - 1} \right)_{kl}  \right]\), which is known by the induction assumption. In order to achieve this, it is necessary to explicitly highlight the existence of such an expression in index notation. The last \(\frac{\partial }{\partial \Lambda} \Lambda\) can be easily separated from \(\left(\mathcal{E}_{m + 1}  \right)_{ij}\) as well as the first \(\Lambda \frac{\partial }{\partial \Lambda}\) from \(\left( \mathcal{E}_{m} \right)_{kl}\). This is possible due to the defining properties of these operators. We obtain
\begin{dmath}
\left( \mathcal{E}_{m + 1} \right)_{ij} \left( \mathcal{E}_m \right)_{kl} = \left(\mathcal{E}_m\right) 
_{ip} \left( \frac{\partial }{\partial \Lambda} \Lambda \right)_{pj}\left( \Lambda\frac{\partial }{\partial \Lambda}  \right)_{kq}\left(\mathcal{E}_{m - 1}\right)_{ql}. 
\end{dmath}
It is possible to commute \(\frac{\partial }{\partial \Lambda} \Lambda\) and \(\Lambda \frac{\partial }{\partial \Lambda}\), both located in the middle, by virtue of \eqref{eq:mc5}
\begin{dmath}
\left( \mathcal{E}_{m + 1} \right)_{ij} \left( \mathcal{E}_m \right)_{kl} = \left(\mathcal{E}_m\right) 
_{ip} \left( \Lambda\frac{\partial }{\partial \Lambda}  \right)_{kq}\left( \frac{\partial }{\partial \Lambda} \Lambda \right)_{pj}\left( \mathcal{E}_{m - 1}\right)_{ql}. 
\end{dmath}
The use of commutators \eqref{eq:mc10} and \eqref{eq:mc11} allows one to move separated \(\Lambda \frac{\partial }{\partial \Lambda}\) and \(\frac{\partial }{\partial \Lambda} \Lambda\) to the boundaries of the expression
\begin{dmath}
\left( \mathcal{E}_{m + 1} \right)_{ij} \left( \mathcal{E}_m \right)_{kl} = \left( \left( \Lambda\frac{\partial }{\partial \Lambda}  \right)_{kq} \left( \mathcal{E}_m\right)_{ip} - \delta_{iq}\left( \mathcal{E}_m\right)_{kp} \right)\left( \left( \mathcal{E}_{m - 1}\right)_{ql} \left( \frac{\partial }{\partial \Lambda} \Lambda \right)_{pj} + \delta_{pl}\left( \mathcal{E}_{m - 1}\right)_{qj} \right).  
\end{dmath}
The expansion of this expression reads as follows:
\begin{dmath}
\label{eq:mc13}
\left( \mathcal{E}_{m + 1} \right)_{ij} \left( \mathcal{E}_m \right)_{kl} =  \left( \Lambda\frac{\partial }{\partial \Lambda}  \right)_{kq} \left( \mathcal{E}_m\right)_{ip} \left( \mathcal{E}_{m - 1}\right)_{ql} \left( \frac{\partial }{\partial \Lambda} \Lambda \right)_{pj} + \left( \Lambda\frac{\partial }{\partial \Lambda}  \right)_{kq} \left( \mathcal{E}_m\right)_{il}\left( \mathcal{E}_{m - 1}\right)_{qj} - \left( \mathcal{E}_m\right)_{kp} \left( \mathcal{E}_{m - 1}\right)_{il} \left( \frac{\partial }{\partial \Lambda} \Lambda \right)_{pj} - \left( \mathcal{E}_m\right)_{kl} \left( \mathcal{E}_{m -  1}\right)_{ij}  . 
\end{dmath}
The assumption of induction
\begin{dmath}
\left[ \left( \mathcal{E}_{m} \right)_{ij}, \left( \mathcal{E}_{m - 1} \right)_{kl} \right] = \left(\mathcal{E}_{m - 1}\right)_{kj} \left(\mathcal{E}_{m - 1}\right)_{il}  
\end{dmath}
can be used to commute the \(\mathcal{E}\)-s in the first two terms of \eqref{eq:mc13}
\begin{dmath}
\left( \mathcal{E}_{m + 1} \right)_{ij} \left( \mathcal{E}_m \right)_{kl} =  \left( \Lambda\frac{\partial }{\partial \Lambda}  \right)_{kq} \left(  \left( \mathcal{E}_{m - 1}\right)_{ql} \left( \mathcal{E}_m\right)_{ip} + \left( \mathcal{E}_{m - 1}\right)_{qp} \left( \mathcal{E}_{m - 1}\right)_{il}\right) \left( \frac{\partial }{\partial \Lambda} \Lambda \right)_{pj} + \left( \Lambda\frac{\partial }{\partial \Lambda}  \right)_{kq} \left( \left( \mathcal{E}_{m - 1}\right)_{qj}\left( \mathcal{E}_m\right)_{il} +  \left( \mathcal{E}_{m - 1}\right)_{ql}\left( \mathcal{E}_{m - 1}\right)_{ij}\right) - \left( \mathcal{E}_m\right)_{kp} \left( \mathcal{E}_{m - 1}\right)_{il} \left( \frac{\partial }{\partial \Lambda} \Lambda \right)_{pj} - \left( \mathcal{E}_m\right)_{kl} \left( \mathcal{E}_{m -  1}\right)_{ij} . 
\end{dmath}
Expanding this expression and contracting the \(\mathcal{E}_m\)-s with \(\frac{\partial }{\partial \Lambda} \Lambda\) or \(\Lambda \frac{\partial }{\partial \Lambda}\) where possible one can get
\begin{dmath}
\left( \mathcal{E}_{m + 1} \right)_{ij} \left( \mathcal{E}_m \right)_{kl} =    \left( \mathcal{E}_{m}\right)_{kl} \left( \mathcal{E}_{m + 1}\right)_{ij} + \left( \mathcal{E}_{m}\right)_{kp} \left( \mathcal{E}_{m - 1}\right)_{il} \left( \frac{\partial }{\partial \Lambda} \Lambda \right)_{pj} + \left( \mathcal{E}_{m}\right)_{kj} \left( \mathcal{E}_m\right)_{il} + \left( \mathcal{E}_{m}\right)_{kl} \left( \mathcal{E}_{m - 1}\right)_{ij} - \left( \mathcal{E}_{m}\right)_{kp} \left( \mathcal{E}_{m - 1}\right)_{il} \left( \frac{\partial }{\partial \Lambda} \Lambda \right)_{pj} - \left( \mathcal{E}_{m}\right)_{kl} \left( \mathcal{E}_{m -  1}\right)_{ij}. 
\end{dmath}
The result is
\begin{dmath}
\left[ \left( \mathcal{E}_{m + 1} \right)_{ij}, \left( \mathcal{E}_m \right)_{kl}  \right] = \left( \mathcal{E}_{m}\right)_{kj} \left( \mathcal{E}_m\right)_{il},  
\end{dmath}
as expected. The subsequent step is to compute the commutator
\begin{dmath}
\left[ E_{m + 1}, H_n^{(m)} \right] = E_{m + 1} H_n^{(m)} -  H_n^{(m)}E_{m + 1}, 
\end{dmath}
the first term of which reads as
\begin{dmath}
E_{m + 1} H_n^{(m)} = \tr \mathcal{E}_{m + 1} \tr \mathcal{E}_m^n = \left( \mathcal{E}_{m + 1} \right)_{ii} \left( \mathcal{E}_m^n \right)_{jj} = \left( \mathcal{E}_{m + 1} \right)_{ii} \delta_{j_1 j_{n + 1}}\prod_{l = 1}^{n}\left( \mathcal{E}_{m} \right)_{j_l j_{l + 1}}.  
\end{dmath}
Then, \(\mathcal{E}_{m + 1}\) in this expression should be successively commuted with all the \(\mathcal{E}_m\)-s, resulting in
\begin{dmath}
E_{m + 1} H_n^{(m)} =  \delta_{j_1 j_{n + 1}}\left( \left( \mathcal{E}_{m} \right)_{j_1 j_{2}}\left( \mathcal{E}_{m + 1} \right)_{ii} + \left( \mathcal{E}_{m} \right)_{j_1 i}\left( \mathcal{E}_{m} \right)_{i j_2} \right)\prod_{l = 2}^{n}\left( \mathcal{E}_{m} \right)_{j_l j_{l + 1}} =  \delta_{j_1 j_{n + 1}} \left( \mathcal{E}_{m} \right)_{j_1 j_{2}}\left( \mathcal{E}_{m + 1} \right)_{ii}\prod_{l = 2}^{n}\left( \mathcal{E}_{m} \right)_{j_l j_{l + 1}} +\tr \mathcal{E}_m^{n + 1} = \delta_{j_1 j_{n + 1}} \left( \prod_{r = 1}^{k}
\left( \mathcal{E}_{m} \right)_{j_r j_{r + 1}} \right)\left( \mathcal{E}_{m + 1} \right)_{ii}\prod_{l = k + 1}^{n}\left( \mathcal{E}_{m} \right)_{j_l j_{l + 1}} +k H_{n + 1}^{(m)} = H_n^{(m)} E_{m + 1} + n H_{n + 1}^{(m)}.
\end{dmath}
This yields the following result
\begin{dmath}
\left[ E_{m + 1}, H_n^{(m)} \right] = n H_{n + 1}^{(m)}. 
\end{dmath}
\subsection{Proof of the recursion relation for \(H_{-n}^{(-m)}\)}
\label{sec:orga09f6f4}
In a complete analogy with the previous subsection, let us prove the relation
\begin{equation}
\label{eq:214}
n H_{-n - 1}^{(-m)} = \left[ H_{-n}^{(-m)}, F_{m + 1} \right],
\end{equation}
For this time, it would be beneficial to have the following auxiliary identity:
\begin{equation}
\label{eq:215}
\left[ \left( \mathcal{F}_m \right)_{ij}, \left( \mathcal{F}_{m  + 1} \right)_{kl} \right] = \left( \mathcal{F}_m \right)_{il} \left( \mathcal{F}_m \right)_{kj}.
\end{equation}
One can prove such an identity by induction with the base
\begin{dmath}
\left[ \left( \frac{\partial }{\partial \Lambda}  \right)_{ij}, \left( \frac{\partial }{\partial \Lambda} \Lambda \frac{\partial }{\partial \Lambda}  \right)_{kl} \right] = \left[ \left( \frac{\partial }{\partial \Lambda}  \right)_{ij}, \left( \frac{\partial }{\partial \Lambda} \right)_{kn}\left(  \Lambda \frac{\partial }{\partial \Lambda}  \right)_{nl} \right] = \left( \frac{\partial }{\partial \Lambda} \right)_{kn}\left[ \left( \frac{\partial }{\partial \Lambda}  \right)_{ij}, \left(  \Lambda \frac{\partial }{\partial \Lambda}  \right)_{nl} \right] = \left( \frac{\partial }{\partial \Lambda}  \right)_{kn}\delta_{jn} \left( \frac{\partial }{\partial \Lambda}  \right)_{il} = \left( \frac{\partial }{\partial \Lambda}  \right)_{kj} \left( \frac{\partial }{\partial \Lambda}  \right)_{il}
\end{dmath}
In light of the induction assumption
\begin{dmath}
\left[ \left( \mathcal{F}_{m - 1} \right)_{ij}, \left( \mathcal{F}_{m} \right)_{kl} \right] = \left( \mathcal{F}_{m - 1} \right)_{il} \left( \mathcal{F}_{m - 1} \right)_{kj} 
\end{dmath}
let us expand the commutator
\begin{equation}
\label{eq:216}
\left[ \left( \mathcal{F}_m \right)_{ij}, \left( \mathcal{F}_{m  + 1} \right)_{kl} \right].
\end{equation}
The first term of such a commutator can be expressed in the following manner:
\begin{dmath}
\left( \mathcal{F}_m \right)_{ij} \left( \mathcal{F}_{m + 1} \right)_{kl} = \left( \mathcal{F}_{m - 1} \right)_{ip} \left( \Lambda \frac{\partial }{\partial \Lambda}  \right)_{pj} \left( \frac{\partial }{\partial \Lambda} \Lambda \right)_{kq} \left( \mathcal{F}_{m} \right)_{ql} = \left( \mathcal{F}_{m - 1} \right)_{ip} \left( \frac{\partial }{\partial \Lambda} \Lambda \right)_{kq} \left( \Lambda \frac{\partial }{\partial \Lambda}  \right)_{pj}  \left( \mathcal{F}_{m} \right)_{ql}.
\end{dmath}
Next, commutators of the \(\mathcal{F}\)-s with \(\frac{\partial }{\partial \Lambda} \Lambda\) and \(\Lambda \frac{\partial }{\partial \Lambda}\) are used to obtain
\begin{dmath}
\left( \mathcal{F}_m \right)_{ij} \left( \mathcal{F}_{m + 1} \right)_{kl}  = \left( \left( \frac{\partial }{\partial \Lambda} \Lambda \right)_{kq}\left( \mathcal{F}_{m - 1} \right)_{ip} + \delta_{iq}\left( \mathcal{F}_{m - 1} \right)_{kp}\right)\left( \left( \mathcal{F}_{m} \right)_{ql} \left( \Lambda \frac{\partial }{\partial \Lambda}  \right)_{pj} - \delta_{lp}\left( \mathcal{F}_m \right)_{qj}\right)  .
\end{dmath}
Expansion of this expression reads as follows:
\begin{dmath}
\left( \mathcal{F}_m \right)_{ij} \left( \mathcal{F}_{m + 1} \right)_{kl}  =  \left( \frac{\partial }{\partial \Lambda} \Lambda \right)_{kq}\left( \mathcal{F}_{m - 1} \right)_{ip}  \left( \mathcal{F}_{m} \right)_{ql} \left( \Lambda \frac{\partial }{\partial \Lambda}  \right)_{pj} -  \left( \frac{\partial }{\partial \Lambda} \Lambda \right)_{kq}\left( \mathcal{F}_{m - 1} \right)_{il}  \left( \mathcal{F}_m \right)_{qj} +   \left( \mathcal{F}_{m - 1} \right)_{kp} \left( \mathcal{F}_{m} \right)_{il} \left( \Lambda \frac{\partial }{\partial \Lambda}  \right)_{pj} -   \left( \mathcal{F}_{m - 1} \right)_{kl}  \left( \mathcal{F}_m \right)_{ij}.
\end{dmath}
Assumption of the induction applied to commute \(\mathcal{F}_{m -1}\) and \(\mathcal{F}_m\) in the first and third terms gives
\begin{dmath}
\left( \mathcal{F}_m \right)_{ij} \left( \mathcal{F}_{m + 1} \right)_{kl}  =  \left( \frac{\partial }{\partial \Lambda} \Lambda \right)_{kq}\left( \left( \mathcal{F}_{m} \right)_{ql}\left( \mathcal{F}_{m - 1} \right)_{ip} + \left( \mathcal{F}_{m - 1} \right)_{il}\left( \mathcal{F}_{m - 1} \right)_{qp}  \right) \left( \Lambda \frac{\partial }{\partial \Lambda}  \right)_{pj} -  \left( \frac{\partial }{\partial \Lambda} \Lambda \right)_{kq}\left( \mathcal{F}_{m - 1} \right)_{il}  \left( \mathcal{F}_m \right)_{qj} +   \left( \left( \mathcal{F}_{m} \right)_{il}\left( \mathcal{F}_{m - 1} \right)_{kp} + \left( \mathcal{F}_{m - 1} \right)_{kl}\left( \mathcal{F}_{m - 1} \right)_{ip} \right) \left( \Lambda \frac{\partial }{\partial \Lambda}  \right)_{pj} -   \left( \mathcal{F}_{m - 1} \right)_{kl}  \left( \mathcal{F}_m \right)_{ij}.
\end{dmath}
The desired relation can be obtained by using the property
\begin{equation}
\label{eq:217}
\mathcal{F}_{m + 1} = \left( \frac{\partial }{\partial \Lambda} \Lambda \right) \mathcal{F}_{m } = \mathcal{F}_m \left( \Lambda \frac{\partial }{\partial \Lambda}  \right)
\end{equation}
and expanding everything in the penultimate equation. It is now necessary to address the commutator:
\begin{dmath}
\left[ H_{-n}^{(-m)}, F_{m + 1}\right] = H_{-n}^{(-m)} F_{m + 1} - F_{m + 1} H_{-n}^{(-m)}. 
\end{dmath}
The first term of the expression is actually
\begin{dmath}
H_{-n}^{(-m)} F_{m + 1} = \tr \mathcal{F}_m^n \tr \mathcal{F}_{m + 1} = \left( \mathcal{F}_m^n \right)_{ii} \left( \mathcal{F}_{m + 1} \right)_{jj} = \left( \prod_{l = 1}^{n} \left( \mathcal{F}_m \right)_{i_l i_{l + 1}}
 \right) \delta_{i_1 i_{n + 1}} \left( \mathcal{F}_{m + 1} \right)_{jj}.
\end{dmath}
By successive commutations of \(\mathcal{F}_{m + 1}\) with all the rest \(\mathcal{F}_m\)-s, one can obtain the final result:
\begin{dmath}
H_{-n}^{(-m)} F_{m + 1} = \left( \prod_{l = 1}^{n - 1} \left( \mathcal{F}_m \right)_{i_l i_{l + 1}}
 \right)  \left( \left( \mathcal{F}_{m + 1} \right)_{jj} \left( \mathcal{F}_m \right)_{i_n i_{n + 1}} +  \left( \mathcal{F}_m \right)_{i_n j} \left( \mathcal{F}_m \right)_{j i_{n + 1}}\right)\delta_{i_1 i_{n + 1}} = \left( \prod_{l = 1}^{n - 1} \left( \mathcal{F}_m \right)_{i_l i_{l + 1}}
 \right)   \left( \mathcal{F}_{m + 1} \right)_{jj} \left( \mathcal{F}_m \right)_{i_n i_{n + 1}}\delta_{i_1 i_{n + 1}} +  \tr \mathcal{F}_m^{n + 1} = F_{m + 1} H_{-n}^{(-m)} + n H_{-n - 1}^{(-m)}.
\end{dmath}
\section{Vertical ray and corresponding \(\widetilde{W}\) algebras}
\label{sec:org23a3d4a}
This section is devoted to a detailed discussion of the vertical ray of the \(W_{1+\infty}\) in the context of the broader theory. The first part concerns its relation to the Hurwitz numbers with completed cycles, while the second one is dedicated to the corresponding \(\widetilde{W}\) algebras.
\subsection{Relation to completed cycles}
\label{sec:org77b6703}
Operators of zero grading can be obtained by commuting \(F_i\) (of grading \(-1\)) with \(E_j\) (of grading 1). A detailed discussion of how to compute such commutators and many others can be found in section \ref{sec:org6fde181}, the conclusion of these computations is that our desired commutators are given via \eqref{eq:213}
\begin{equation}
\label{eq:223}
 H_{0}^{(m)} = \sum_{n = 0}^{ m} \tr\left( \left( \frac{\partial }{\partial \Lambda} \Lambda \right)^{m - n}  \left( \Lambda \frac{\partial }{\partial \Lambda} \right)^{n}  \right). 
\end{equation}
It can be shown that these operators can be expanded only in terms of \(\Lambda \frac{\partial }{\partial \Lambda}\) while acting on analytic functions of \(\Lambda\)
\begin{equation}
\label{eq:219}
f(\Lambda) = \sum_{n = 0}^{\infty} \Lambda^n f_n \left( \mathbf{p} \right).
\end{equation}
This statement is supported by the following observation. Let us consider  the commutator of \(\Lambda\) and \(\frac{\partial }{\partial \Lambda}\) over the space of such functions:
\begin{dmath}
\label{eq:220}
\left[ \frac{\partial }{\partial \Lambda}, \Lambda  \right] f(\Lambda) = \frac{\partial }{\partial \Lambda} \Lambda f(\Lambda) - \Lambda \frac{\partial }{\partial \Lambda} f(\Lambda) = \frac{\partial }{\partial \Lambda} \sum_{n = 0}^{\infty}\Lambda^{n + 1} f_n(\mathbf{p}) - \Lambda \frac{\partial }{\partial \Lambda}  \sum_{n = 0}^{\infty} \Lambda^n f_n (\mathbf{p}) = \sum_{n = 0}^{\infty} \left( f_n (\mathbf{p}) \frac{\partial }{\partial \Lambda} \Lambda^{n + 1} + \Lambda^{n + 1} \frac{\partial }{\partial \Lambda}  f_n(\mathbf{p}) - f_n(\mathbf{p}) \Lambda \frac{\partial }{\partial \Lambda} \Lambda^n - \Lambda^{n + 1} \frac{\partial }{\partial \Lambda} f_n(\mathbf{p}) \right) = \sum_{n = 0}^{\infty} f_n(\mathbf{p})\left( \sum_{k = 0}^{n}\Lambda^{n-k}p_k  -  \sum_{k = 0}^{n - 1} \Lambda^{n-k} p_k  \right) = \tr \left( I f(\Lambda) \right).
\end{dmath}
This result motivates the introduction of the trace operator \(\trh\), which acts on everything to the right of it. In these terms
\begin{equation}
\label{eq:221}
\left[ \frac{\partial }{\partial \Lambda} ,\Lambda \right] = \trh I.
\end{equation}
It is important because it allows for the expansion of all the \(\frac{\partial }{\partial \Lambda}\Lambda\) in terms of \(\Lambda \frac{\partial }{\partial \Lambda}\)
\begin{equation}
\label{eq:222}
\frac{\partial }{\partial \Lambda} \Lambda = \Lambda \frac{\partial }{\partial \Lambda} + \trh I
\end{equation}
and eq. \eqref{eq:223} can be rewritten as
\begin{equation}
\label{eq:224}
H_{0}^{(m)} = \sum_{n = 0}^{ m} \trh \left( \Lambda \frac{\partial }{\partial \Lambda} + \trh I \right)^{m - n}  \left( \Lambda \frac{\partial }{\partial \Lambda} \right)^{n} . 
\end{equation}
This expression means exactly that \(H_0^{(m)}\) can be expressed only trough \(\Lambda \frac{\partial }{\partial \Lambda}\) operators. For the sake of clarity, let us list the first few \(H_{0}^{(m)}\)-s, obtained in this fashion:
\begin{equation}
\label{eq:225}
\begin{aligned}
H_0^{(0)} & = N,\\
H_0^{(1)} & = 2 \tr \left( \Lambda \frac{\partial }{\partial \Lambda}  \right) + N^2,\\
H_0^{(2)} & = 3 \tr \left( \Lambda \frac{\partial }{\partial \Lambda}  \right)^2 + 3N \tr\left( \Lambda \frac{\partial }{\partial \Lambda}  \right) + N^3 = 6 W_0,\\
H_0^{(3)} & = 4 \tr\left( \Lambda \frac{\partial }{\partial \Lambda}  \right)^3 + 4 N \tr\left( \Lambda \frac{\partial }{\partial \Lambda}  \right)^2 + 4N^2 \tr\left( \Lambda \frac{\partial }{\partial \Lambda}  \right) + 2  \tr\nolimits^2 \left( \Lambda \frac{\partial }{\partial \Lambda}  \right) . 
\end{aligned}
\end{equation}
The first and most obvious observation is that these operators can be expanded in terms of the generalized cut-and-join operators \cite{mironov-2011-compl-set}
\begin{equation}
\label{eq:218}
W_{\lambda} = \normord{\prod_{i = 1}^{\ell(\lambda)}\tr \left( \Lambda \frac{\partial }{\partial \Lambda}  \right)^{\lambda_i}
},
\end{equation}
where \(\lambda\) is a partition \(\lambda = \left\{ \lambda_1, \lambda_2,\ldots \right\}\), \(\lambda_1 \ge \lambda_2\ge\ldots\), and the normal ordering means that all the derivatives don't act on \(\Lambda\)-s inside \(W_{\lambda}\). This is a self-evident conclusion, given that the set of cut-and-join operators provides a basis for operators \(\mathcal{O}\left( \Lambda \frac{\partial }{\partial \Lambda}  \right)\). The cut-and-join operator \(W_0\) \cite{goulden-2000-number-ramif,mironov-2011-compl-set} is known to be  the trigonometric Calogero-Sutherland Hamiltonian \cite{morozov-2010-new-old}. It can be expressed through the generalized operators:
\begin{equation}
\label{eq:227}
W_0 = W_{\left\{ 2 \right\}} - N W_{\left\{ 1 \right\}} - \frac{N^3}{6}.
\end{equation}
Subsequently, let us see what interesting can be found while considering the corresponding \(W\)-representation for the \(H_0^{(m)}\) operators (see a discussion in \cite{mironov-2023-spect-curves,mulase-2013-spect-curve,shadrin-2014-equiv-elsv,kramer-2019-towar-orbif})
\begin{dmath}
\label{eq:226}
Z_0^{(m)} = e^{\frac{1}{m} H_0^{(m)}} e^{\sum_{n}^{} \frac{g_n p_n}{n}}.
\end{dmath}
The Cauchy identity \cite{macdonald-1998-symmet-funct} allows one to expand the exponent into a sum over partitions \(\lambda\) of products of Schur polynomials \(S_{\lambda}\)
\begin{equation}
\label{eq:228}
e^{\sum_{n}^{} \frac{g_n p_n}{n}} = \sum_{\lambda}^{} S_{\lambda}(\mathbf{p}) S_{\lambda} ( \mathbf{g}).
\end{equation}
To move further one should notice, at this time, a not so trivial fact: the Schur polynomials are eigenfunctions of \(H_0^{(m)}\)
\begin{equation}
\label{eq:229}
H_0^{(m)} S_{\lambda}(\mathbf{p}) = \mathcal{E}_m (\lambda) S_{\lambda} (\mathbf{p})
\end{equation}
with eigenvalues
\begin{equation}
\label{eq:230}
\mathcal{E}_n (\lambda) = \sum_{j \ge 0}^{} \frac{1}{4^j} \binom{n}{2j + 1} \tilde{C} _{n -2j -1}(\lambda) + N^n,
\end{equation}
where
\begin{equation}
\label{eq:232}
\tilde{C}_k (\lambda) = \sum_{i}^{} \left( \lambda_i -i + \frac{1}{2} + N \right)^k - \left( -i + \frac{1}{2} + N \right)^k - N^k
\end{equation}
are linearly related to the celebrated shifted symmetric power sums
\begin{equation}
\label{eq:233}
C_k (\lambda) = \sum_{i}^{} \left( \lambda_{i} -i + \frac{1}{2} \right)^k - \left( -i + \frac{1}{2} \right)^k.
\end{equation}
This is an empirical fact, although it is strongly supported by computational evidence. What are the implications of this result? Now it is evident that
\begin{equation}
\label{eq:234}
Z_0^{(m)} = \sum_{\lambda}^{} S_{\lambda}(\mathbf{p}) S_{\lambda} (\mathbf{g}) e^{\frac{1}{m}\mathcal{E}_m(\lambda)},
\end{equation}
where \(\mathcal{E}_m(\lambda)\)-s are linearly related to shifted symmetric power sums \(C_{k}(\lambda)\). This allows me to conclude that \(Z_0^{(m)}\) is a KP \(\tau\)-function \cite{kharchev-1995-gener-kazak,orlov-2001-hyper-solut,orlov-2006-hyper-funct,okounkov-2006-gromov-witten,2008-combin-hurwit,alexandrov-2012-integ-hurwit}. Moreover, it can be demonstrated that \(Z_0^{(m)}\) is a hypergeometric \(\tau\)-function \cite{orlov-2001-hyper-solut,orlov-2006-hyper-funct}. Two sets of variables, \(p_k\) and \(g_k\), are on equal footing in \(Z_0^{(m)}\), thus it is a KP \(\tau\)-function with respect to both sets of variables. The dependence on the Toda zeroth time necessitates further specification \cite{takasaki-1984-initial-value}, resulting in the Toda lattice hierarchy \cite{ueno-1984-toda-lattic-hierar}. One may also pursue an alternative approach, whereby it can be demonstrated that \(Z_0^{(m)}\) is a KP \(\tau\)-function as soon as \(H_0^{(m)}\) is an element of the \(W_{1 + \infty}\) algebra \cite{orlov-1988,orlov-1997-p-algeb,takasaki-1999-quasic-limit,takasaki-1993-quasi-class}.

A few words about a non-completed cycles are in order. The eigenfunctions of the generalized cut-and-join operators \(W_{\{k\}}\) are, once again, the Schur functions
\begin{equation}
\label{eq:235}
W_{\{k\}} S_{\lambda} (\mathbf{p}) = \phi_{\lambda} (\{k\}) S_{\lambda}(\mathbf{p}),
\end{equation}
with an eigenvalue \(\phi_{\lambda}(\{k\})\) proportional \cite{mironov-2011-compl-set,ivanov-2001,mironov-2012-algeb-differ} to the value of the character of the symmetric group \(S_n\), \(n = |\lambda|\), in the representation \(\lambda\) on the element with the only non-unit cycle of length \(k\) \cite{fulton-1997-young-tableaux}. By employing the same methodology as previously the following identity is derived
\begin{equation}
\label{eq:236}
e^{W_{\{k\}}} \cdot 1 = \sum_{\lambda}^{} S_{\lambda} (\mathbf{p})  S_{\lambda} (\mathbf{g}) e^{\phi_{\lambda}(\{k\})}.
\end{equation}
It turns out that for \(k >2\) only non-linear combinations of \(\phi_{\lambda}(\{k\})\) give the shifted symmetric sums \(C_n(\lambda)\) \cite{mironov-2011-compl-set,mironov-2021-connec-between}. Such combinations are referred to as \emph{completed cycles} \cite{okounkov-2006-gromov-witten,2008-combin-hurwit}. The completed cycles have received considerable attention during the last years in the context of enumerative geometry (see, for instance \cite{chiodo-2008-towar-enumer,alexandrov-2017-fermion-approac,alexandrov-2016-ramif-hurwit}). In particular, they feature in the celebrated Zvonkine's conjecture \cite{zvonkine-2006-prepr}, which was recently proved in \cite{dunin-barkowski-2023-loop-equat}. The enumerative geometric meaning of  the non-vertical \(W_{1 + \infty}\) rays remains unclear, and it represents a promising area for future investigation.
\subsection{Corresponding \(\widetilde{W}\) algebra}
\label{sec:org8a91615}
It is now relatively straightforward to derive the \(\widetilde{W}\) algebra for the Hamiltonians
\begin{equation}
\label{eq:237}
 H_{0}^{(m)} = \sum_{n = 0}^{ m} \tr\left( \left( \frac{\partial }{\partial \Lambda} \Lambda \right)^{m - n}  \left( \Lambda \frac{\partial }{\partial \Lambda} \right)^{n}  \right).  
\end{equation}
The main non-recursive definition for this case will be as follows:
\begin{equation}
\label{eq:238}
\sum_{n = 0}^{ m} \left( \frac{\partial }{\partial \Lambda} \Lambda \right)^{m - n}  \left( \Lambda \frac{\partial }{\partial \Lambda} \right)^{n} f(\mathbf{p}) = \sum_{k}^{} \Lambda^k \widetilde{W}_k^{(m, 0)}f(\mathbf{p}).
\end{equation}
This definition can be naturally expanded into the auxiliary ones
\begin{equation}
\label{eq:239}
\left( \frac{\partial }{\partial \Lambda} \Lambda \right)^{m - n}  \left( \Lambda \frac{\partial }{\partial \Lambda} \right)^{n}f(\mathbf{p})  = \sum_{k}^{} \Lambda^k \widetilde{W}_k^{(m, 0|n)}f(\mathbf{p}) \qquad \text{for } m \ge n,
\end{equation}
\begin{equation}
 \left( \Lambda \frac{\partial }{\partial \Lambda} \right)^{m}f(\mathbf{p})  = \sum_{k}^{} \Lambda^k \widetilde{W}_k^{(m, 0|n)}f(\mathbf{p}) \qquad \text{otherwise}. 
\end{equation}
The immediate consequence of such a definition
\begin{equation}
\label{eq:240}
\widetilde{W}_k^{(m,0|n)} = 0 \qquad \text{for} \qquad k<0.
\end{equation}
The base of the corresponding recursion, as always,
\begin{equation}
\label{eq:241}
\widetilde{W}_k^{(0,0|n)} = \delta_{k,0}.
\end{equation}
Next, it is instructive to consider the case of \(m < n\)
\begin{equation}
\label{eq:242}
\left( \Lambda \frac{\partial }{\partial \Lambda}  \right)^m f(\mathbf{p})= \sum_{k}^{} \Lambda^k \widetilde{W}_k^{(m,0|n)} f(\mathbf{p}).  
\end{equation}
This is nothing but the definition of \(\widetilde{W}_k^{(m, -)}\) operators. For \(m < n\), the recursion procedure is as follows:
\begin{equation}
\label{eq:243}
\widetilde{W}_k^{(m + 1, 0|n)} = \sum_{r \ge 0}^{} p_r \widetilde{W}_{k + r}^{(m, 0|n)} +
\sum_{r = 1}^{k }r \frac{\partial }{\partial p_r}  \widetilde{W}_{k - r}^{(m, 0|n)} \qquad \text{for } \qquad k \ge 1,
\end{equation}
\begin{equation}
\label{eq:244}
\widetilde{W}_k^{(m + 1, 0 |n )} = 0, \qquad \text{otherwise.}
\end{equation}
For \(m \ge n\) we have
\begin{dmath}
\label{eq:hi}
\sum_{k \ge 0}^{} \Lambda^k \widetilde{W}_k^{(m + 1, 0\mid  n)} f(\mathbf{p}) = \left( \frac{\partial }{\partial \Lambda} \Lambda \right)^{m - n + 1}  \left( \Lambda \frac{\partial }{\partial \Lambda} \right)^{n}f(\mathbf{p})  = \left( \frac{\partial }{\partial \Lambda} \Lambda \right) \left( \frac{\partial }{\partial \Lambda} \Lambda \right)^{m - n}  \left( \Lambda \frac{\partial }{\partial \Lambda} \right)^{n}f(\mathbf{p})= \left( \frac{\partial }{\partial \Lambda} \Lambda \right) \sum_{k \ge 0}^{} \Lambda^k \widetilde{W}_k^{(m, 0|n)}f(\mathbf{p}).
\end{dmath}
This is exactly the recursion of \(\widetilde{W}_k^{(m, +)}\) operators. Consequently, for \(m \ge n\)
\begin{equation}
\widetilde{W}_k^{(m + 1, 0|n)} = \sum_{r \ge 0}^{} p_r \widetilde{W}_{k + r}^{(m, 0|n)} +
\sum_{r = 1}^{k }r \frac{\partial }{\partial p_r}  \widetilde{W}_{k - r}^{(m, 0|n)} \qquad \text{for } \qquad k \ge 0,
\end{equation}
\begin{equation}
\widetilde{W}_k^{(m + 1, 0 |n )} = 0, \qquad \text{otherwise.}
\end{equation}
The summary of the aforementioned procedure
\begin{subequations}
\label{eq:251}
\begin{empheq}[box=\fbox]{gather}
\widetilde{W}_k^{(m, 0)} = \sum_{n = 0}^{m}\widetilde{W}_k^{(m, 0 | n)}, \\
\begin{aligned}
\widetilde{W}_{k}^{(m + 1, 0 | n)}  = \sum_{ \ge 0}^{}p_r \widetilde{W}_{k + r}^{(-m, -n |-l)}
+ \sum_{r = 1}^{k} r \frac{\partial }{\partial p_r}  &\widetilde{W}_{k - r}^{(m, 0 |n)} \\
\quad &\text{for}\quad k > - \theta (m - n),
\end{aligned} \\
\widetilde{W}_k^{(0, 0 | n)} = \delta_{k, 0}, \\
\widetilde{W}_k^{(m, 0 | n)} = 0 \qquad \text{otherwise,} 
\end{empheq}
\end{subequations}
where \(\theta(x)\) is the Heaviside theta, defined as
\begin{equation}
\label{eq:245}
\theta(x) = \begin{cases}
  1, & x \ge 0,\\
0, & x<0.
\end{cases}
\end{equation}
\section{Conclusions}
\label{sec:org5104969}
Finally, we will present the primary outcomes presented in the study, as follows:
\begin{itemize}
\item The non-recursive definitions of the generalized \(\widetilde{W}\) algebras are formulated for the positive \eqref{eq:161}, negative \eqref{eq:163}, and zero \eqref{eq:238} grading \(W_{1+ \infty}\) Hamiltonians.
\item From this, the corresponding recursive definitions \eqref{eq:185}, \eqref{eq:200}, \eqref{eq:251} are derived.
\item The Ward identities for the generalized WLZZ partition functions \(Z_n^{(m)}\) \eqref{eq:64} are also obtained.
\item Some useful matrix calculus identities are proved or derived.
\end{itemize}
Possible directions for future research:
\begin{itemize}
\item \(\beta\)-deformation of this generalized \(\widetilde{W}\) algebras and further study of \(\beta\)-deformed WLZZ models (for recent advances in this field see e.g. \cite{mironov-2024-two-realiz,mironov-2024-wlzz-model,morozov-2024-charac-expan}).
\item Finding the enumerative geometric meaning of the non-vertical rays (in analogy to the correspondence between the vertical ray and the Hurwitz numbers with completed cycles).
\item Generalizing the concept of the \(\widetilde{W}\) algebras even more, to correspond to an arbitrary operator \(\mathcal{O} \left( \Lambda, \frac{\partial }{\partial \Lambda}  \right)\), not only \(H_n^{(m)}\)-s.
\item Formulating further a matrix analog of the \(w_{\infty}\) algebra, finding commutation relations for the elements of such an algebra.
\item Rigorous proof of conjectured eq. \eqref{eq:64} is also a very important problem.
\end{itemize}
\section*{Acknowledgments}
\label{sec:org3c189ca}
I am immensely grateful to A. Mironov and A. Popolitov for enlightening discussions. This work was supported by the Russian Science Foundation (Grant No.20-71-10073).

\printbibliography
\appendix
\section{Computation of \(\widetilde{W}_k^{(1)}\) and \(\widetilde{W}_k^{(2)}\)}
\label{sec:org1a24712}
For \(k \ge 0\) we have 
\begin{equation}
\label{eq:93}
\widetilde{W}_k^{(1)} =\sum_{m \ge 0}^{} p_m \delta_{k + m, 0} + \sum_{m = 1}^{k} m \frac{\partial }{\partial p_m} \delta_{k - m, 0},   
\end{equation}
The process of computing the restricted sums containing the Kronecker deltas can be more involved than a simple interchange of indices in the expression. The general recipe for such sums is as follows:
\begin{equation}
\label{eq:116}
\sum_{i \in I}^{}f_i \delta_{i,j} = f_j \sum_{i \in I}^{}\delta_{i,j}, 
\end{equation}
What is written here is that in comparison to an unrestricted sum, a non-zero value can be obtained only if the summation range contains an index that should be substituted. Let us now turn our attention to the first sum in \eqref{eq:93} 
\begin{equation}
\label{eq:114}
\sum_{m \ge 0 }^{}p_m \delta_{k+m,0} = p_{-k} \sum_{m \ge 0}^{} \delta_{k + m, 0} = p_{-k} \sum_{m \le 0}^{} \delta_{k, m}, 
\end{equation}
It is necessary to recall the restriction \(k \ge 0\) and apply it to the expression above. All the Kronecker deltas, with the exception of  \(\delta_{k,0}\), will automatically vanish for non-negative \(k\)
\begin{equation}
\label{eq:122}
\sum_{m \ge 0}^{}p_m \delta_{k + m, 0} = p_0 \delta_{k,0} = N \delta_{k,0}.  
\end{equation}
The second term is easier to analyze. The value \(k = m\) is within the summation range, so here \(m\) will be simply substituted by \(k\) as it works in the unrestricted sums
\begin{equation}
\label{eq:118}
\sum_{m = 1}^{k} m \frac{\partial }{\partial p_m} \delta_{k - m, 0} = 
k \frac{\partial }{\partial p_k} . 
\end{equation}
As a result, for \(k \ge 0\)
\begin{equation}
\label{eq:119}
\widetilde{W}_k^{(1)} = N \delta_{k,0} + k \frac{\partial }{\partial p_k}.  
\end{equation}
Next, for \(k \ge -1\) we have
\begin{dmath}
\label{eq:94}
\widetilde{W}_k^{(2)} =  \sum_{m \ge 0}^{} p_m \left( N \delta_{k + m, 0} + (k + m) \frac{\partial }{\partial p_{k + m}}  \right) + \sum_{m = 1}^{k + 1} m \frac{\partial }{\partial p_m} \left( N \delta_{k - m,0} + (k - m) \frac{\partial }{\partial p_{k - m}} \right).  
\end{dmath}
After taking the sums using the described methods
\begin{dmath}
\label{eq:120}
\widetilde{W}_k^{(2)} = p_{-k} N \sum_{m \le 0}^{} \delta_{k,m} + \sum_{m \ge 0}^{}(k + m)p_m \frac{\partial }{\partial p_{k + m}} + Nk \frac{\partial }{\partial p_k} + \sum_{m = 1}^{k + 1} m (k - m) \frac{\partial ^2}{\partial p_m \partial p_{k - m}}.   
\end{dmath}
In this context, the restriction to \(k \ge -1\) results in the survival of two deltas out of the entire sum, namely \(\delta_{k,0} + \delta_{k, -1}\). Additionally, the upper bound of the last sum is too high. This is evidenced by the vanishing of the \(k + 1\)-th term due to the derivative \(\frac{\partial }{\partial p_{-1}}\) and the \(k\)-th term due to the multiplier \((k - m)\). Consequently, the sum can be safely taken only up to \(k - 1\). As a result,
\begin{dmath}
\label{eq:123}
\widetilde{W}_k^{(2)} = N p_1 \delta_{k, -1} + N^2 \delta_{k,0} + \sum_{m \ge 0}^{}(k + m)p_m \frac{\partial }{\partial p_{k + m}} + Nk \frac{\partial }{\partial p_k} + \sum_{m = 1}^{k - 1} m (k - m) \frac{\partial ^2}{\partial p_m \partial p_{k - m}}.    
\end{dmath}
It can be seen that the computation of the \(\widetilde{W}_k^{(3)}\) and all the higher \(\widetilde{W}_k^{(n)}\)-s is a conceptually simple task.
\section{Matrix calculus and proof of \eqref{eq:92}}
\label{sec:org8a93151}
The derivative with respect to \(N \times N\) matrix \(\Lambda\) is defined as
\begin{equation}
\label{eq:1}
\left( \frac{\partial }{\partial \Lambda} \right)_{ij} = \frac{\partial }{\partial \Lambda_{ji}}. 
\end{equation}
In this paper the matrix derivatives act on smooth scalar functions of matrix variable \(\Lambda\). These functions can be given as a series in a possibly infinite set of variables \(\mathbf{p} = \left\{ p_0, p_1, p_2, \dots \right\}\), where each \(p_k\) is equal to \(\tr \Lambda^k\). These variables can also can be traces of inverses, but this will be stated explicitly where necessary. In our calculations, we typically expand the action of such derivatives on scalar functions using the chain rule
\begin{equation}
\label{eq:2}
\frac{\partial }{\partial \Lambda} f ( \mathbf{p} ) = \sum_{k \ge 0}^{}
\frac{\partial p_k}{\partial \Lambda} \frac{\partial }{\partial p_k} f ( \mathbf{p} ) 
\end{equation}
and the matrix calculus identity
\begin{equation}
\label{eq:3}
\frac{\partial p_k}{\partial \Lambda} = \frac{\partial }{\partial \Lambda} \tr \Lambda^k 
= k \Lambda^{k - 1}
\end{equation}
to get
\begin{equation}
\label{eq:4}
\frac{\partial }{\partial \Lambda} f(\mathbf{p}) = \sum_{k\ge 1}^{}k \Lambda^{k - 1} 
\frac{\partial }{\partial p_k} f(\mathbf{p}).
\end{equation}
One additional useful matrix calculus identity will be employed until the end of the paper
\begin{equation}
\label{eq:5}
\frac{\partial }{\partial \Lambda} \Lambda^k = \sum_{r = 0}^{k - 1}\Lambda^{k - r - 1} \tr \Lambda^{r} = \sum_{r = 0}^{k - 1}\Lambda^{k - r - 1} p_r, 
\end{equation}
where the matrix derivative and the matrix power here are contracted in a manner analogous to the regular matrix product. It should be noted, however, that this formula is only valid for \(k \ge 0\). The rule of differentiating negative matrix powers differs slightly from the rule for the positive ones.

The final piece of matrix calculus that will be of interest is the expression for the matrix derivative of the determinant. This expression is given by:
\begin{equation}
\label{eq:72}
\frac{\partial }{\partial \Lambda} \det \Lambda = \Lambda^{-1} \det \Lambda. 
\end{equation}
The objective is to prove \eqref{eq:92}. To do this the expression for the derivative by an inverse matrix
\begin{equation}
\label{eq:73}
\frac{\partial }{\partial \Lambda^{-1}}
\end{equation}
should be derived in the first place. In index notation
\begin{equation}
\label{eq:74}
\left(  
\frac{\partial }{\partial\left( \Lambda^{-1} \right)} \right)_{ij}= 
\frac{\partial}{\partial \left( \Lambda^{-1} \right)_{ji}}
\end{equation}
one can use the chain rule
\begin{equation}
\label{eq:75}
 \frac{\partial}{\partial \left( \Lambda^{-1} \right)_{ji}} = 
 \frac{\partial \Lambda_{kl}}{\partial \left( \Lambda^{-1} \right)_{ji}} \frac{\partial}{\partial \Lambda_{kl}}.
\end{equation}
We have faced a new problem of finding the expression
\begin{equation}
\label{eq:76}
\frac{\partial \Lambda_{kl}}{\partial \left( \Lambda^{-1} \right)_{ji}}, 
\end{equation}
which can be solved by differentiating the definition of the inverse matrix
\begin{equation}
\label{eq:77} 
\Lambda \Lambda^{-1} = 1
\end{equation}
with respect to the inverse matrix element
\begin{equation}
\label{eq:78} 
\left. \frac{\partial}{\partial \left( \Lambda^{-1} \right)_{ji}} \right| \Lambda_{km} \left( \Lambda^{-1} \right)_{mn} =\delta_{kn},
\end{equation}
\begin{equation}
\label{eq:79} 
\frac{\partial \Lambda_{km}}{\partial \left( \Lambda^{-1} \right)_{ji}} \left( \Lambda^{-1} \right) _{mn} +
\Lambda_{km}\frac{\partial \left( \Lambda^{-1} \right)_{mn}}{\partial \left( \Lambda^{-1} \right)_{ji}}=0,
\end{equation}
\begin{equation}
\label{eq:80} 
\frac{\partial \Lambda_{km}}{\partial \left( \Lambda^{-1} \right)_{ji}}\left( \Lambda^{-1} \right) _{mn} +
\Lambda_{km} \delta_{mj} \delta_{ni}=0,
\end{equation}
\begin{equation}
\label{eq:81} 
\left. \frac{\partial \Lambda_{km}}{\partial \left( \Lambda^{-1} \right)_{ji}}\left( \Lambda^{-1} \right) _{mn} +
\Lambda_{kj} \delta_{ni}=0 \right| \cdot \Lambda_{nl},
\end{equation}
\begin{equation}
\label{eq:82} 
\frac{\partial \Lambda_{km}}{\partial \left( \Lambda^{-1} \right)_{ji}}\delta_{ml} +
\Lambda_{kj} \Lambda_{il}=0 ,
\end{equation}
\begin{equation}
\label{eq:83} 
\frac{\partial \Lambda_{kl}}{\partial \left( \Lambda^{-1} \right)_{ji}} = -
\Lambda_{kj} \Lambda_{il}.
\end{equation}
Now we have
\begin{equation}
\label{eq:84} 
\frac{\partial}{\partial \left( \Lambda^{-1} \right)_{ji}} = - \Lambda_{kj} \Lambda_{il} \frac{\partial}{\partial \Lambda_{kl}} = -\Lambda_{il} \frac{\partial}{\partial \Lambda_{kl}} \Lambda_{kj} + N\Lambda_{ij},
\end{equation}
so
\begin{equation}
\label{eq:85} 
\frac{\partial}{\partial \Lambda^{-1}} = -\Lambda \frac{\partial }{\partial \Lambda} \Lambda + N \Lambda.
\end{equation}
Armed with such a formula it is not a hard task to prove the desired expression
\begin{dmath}
\label{eq:86}
-\det\nolimits^{-N} \Lambda \frac{\partial }{\partial \Lambda^{-1}} \det\nolimits^N \Lambda =
-\det\nolimits^{-N} \Lambda \left( \frac{\partial }{\partial \Lambda^{-1}} \det\nolimits^{-N} \Lambda^{-1} \right)
-\frac{\partial }{\partial \Lambda^{-1}} = 
N \Lambda + \Lambda \frac{\partial }{\partial \Lambda} \Lambda - N \Lambda = \Lambda \frac{\partial }{\partial \Lambda} \Lambda.
\end{dmath}
\end{document}